%% file: main.tex
\documentclass[sigconf]{acmart}

\setcopyright{none}
\renewcommand\footnotetextcopyrightpermission[1]{}

\newcommand\appendixref{Appendix\xspace}

\iftrue
\newcommand{\yuxin}[1]{\authnote{Yuxin}{#1}}
\newcommand{\zeyu}[1]{\authnote{Zeyu}{#1}}
\newcommand{\dan}[1]{\authnote{Dan}{#1}}
\newcommand{\danfeng}[1]{\authnote{Danfeng}{#1}}
\else
\newcommand{\yuxin}[1]{}
\newcommand{\zeyu}[1]{}
\newcommand{\dan}[1]{}
\newcommand{\danfeng}[1]{}
\fi

\startPage{1}

\usepackage[utf8]{inputenc}
\usepackage[T1]{fontenc}
\usepackage{microtype}
\usepackage{amsmath,amsthm,commath}
\usepackage{mathtools}
\usepackage[linesnumbered,lined,ruled,vlined]{algorithm2e}
\usepackage{paralist}
\usepackage{tablefootnote}
\usepackage{listings}
\usepackage{mathpartir} 
\usepackage{stmaryrd}
\SetSymbolFont{stmry}{bold}{U}{stmry}{m}{n}
\usepackage{xspace}
\usepackage{dsfont}
\usepackage{bbold}
\usepackage{setspace}
\usepackage{balance}
\usepackage{subcaption}
\usepackage{comment}
\usepackage{color}
\usepackage{xr}
\usepackage[normalem]{ulem}
\usepackage{enumitem}
\usepackage{makecell}
\usepackage{footnote}
\usepackage{anyfontsize}
\usepackage{multirow}
\usepackage{colortbl}
\usepackage{declarations}

\newcommand{\authnote}[2]{({\bf \textcolor{blue}{#1}: \em \textcolor{red}{#2}})}


\title{\tool: Automated Program Synthesis for Differential Privacy}

\author{Yuxin Wang}
\affiliation{
  \institution{Pennsylvania State University}
  \city{University Park}
  \state{PA}
  \postcode{16802}
  \country{USA}
}
\email{yxwang@psu.edu}
\author{Zeyu Ding}
\affiliation{
  \institution{Pennsylvania State University}
  \city{University Park}
  \state{PA}
  \postcode{16802}
  \country{USA}
}
\email{zyding@psu.edu}
\author{Yingtai Xiao}
\affiliation{
  \institution{Pennsylvania State University}
  \city{University Park}
  \state{PA}
  \postcode{16802}
  \country{USA}
}
\email{yxx5224@psu.edu}
\author{Daniel Kifer}
\affiliation{
  \institution{Pennsylvania State University}
  \city{University Park}
  \state{PA}
  \postcode{16802}
  \country{USA}
}
\email{dkifer@cse.psu.edu}
\author{Danfeng Zhang}
\affiliation{
  \institution{Pennsylvania State University}
  \city{University Park}
  \state{PA}
  \postcode{16802}
  \country{USA}
}
\email{zhang@cse.psu.edu}

\begin{document}
\fancyhead{} %

\begin{abstract}

Differential privacy has become a de facto standard for releasing data in a privacy-preserving way. Creating a differentially private algorithm is a process that often starts with a noise-free (non-private) algorithm. The designer then decides where to add noise, and how much of it to add. This can be a non-trivial process -- if not done carefully, the algorithm might either violate differential privacy or have low utility.

In this paper, we present \tool, a program synthesizer that takes in non-private code (without any noise) and automatically synthesizes its differentially private version (with carefully calibrated noise). Under the hood, \tool uses novel algorithms to automatically generate a sketch program with candidate locations for noise, and then optimize privacy proof and noise scales simultaneously on the sketch program. Moreover, \tool can synthesize sophisticated mechanisms that adaptively process  queries  until a specified privacy budget is exhausted. When evaluated on standard benchmarks, \tool is able to generate differentially private mechanisms that optimize simple utility functions within 120 seconds. It is also powerful enough to synthesize adaptive privacy mechanisms.

\end{abstract}

\maketitle

\section{Introduction}\label{sec:intro}

\input{sec_introduction}

\section{Background}\label{sec:primilinaries}
\input{sec_preliminaries}

\section{Overview}\label{sec:overview}
\input{sec_overview}

\section{Sketch Generation}
\label{sec:sketch}
\input{sec_sketch}

\section{Implementation and Evaluation}\label{sec:evaluation}
\input{sec_implementation}

\section{Related Work}
\input{sec_related.tex}

\section{Conclusions and Future Work}
\input{sec_conclusion.tex}

\section*{Acknowledgments}
We thank the anonymous CCS reviewers for their insightful feedbacks. This work was supported by NSF Awards CNS-1702760, CNS-1931686 and a gift from Facebook.

\bibliographystyle{ACM-Reference-Format}
\bibliography{diffpriv}
\clearpage
\nobalance
\appendix
\input{sec_appendix.tex}

\end{document}

%% file: sec_introduction.tex
Differential privacy~\cite{dwork06Calibrating} has become a de facto standard for releasing data in a privacy-preserving way. It has been increasingly adopted in industry \cite{rappor,prochlo,applediffp,DingKY17,elasticsensitivity} and the public sector \cite{ashwin08:map,onthemap,Haney:2017:UCF,abowd18kdd}.
Crucial to any differentially private system is a set of \emph{privacy mechanisms}, the building blocks of larger privacy-preserving algorithms. Privacy mechanisms inject randomness into non-private computations in order to ensure privacy protections. However, developing such mechanisms is a daunting task for several reasons. 
\begin{itemize}
\item From the privacy perspective, one must carefully choose exactly where the noise must be injected and how much noise to use. Such decisions are notoriously tricky. For example, the Sparse Vector Technique (SVT) \cite{diffpbook} is designed to return the identities of $N$ queries whose answers are likely to be larger than a public threshold $T$. Lyu et al.~\cite{ninghuisparse} catalog several peer-reviewed yet incorrect variants of SVT. At the source code level, these incorrect variants are very similar to the correct ones, but the tiny differences broke their privacy properties. 
\item From the utility perspective, there are many ways of converting a non-private program into a differentially private one -- each such option could have wildly different utility properties. For instance, in the aforementioned SVT, injecting noise in one specific place (the threshold) allows the mechanism to use much less noise everywhere else (while processing more queries). Aside from different valid choices in noise locations, the mechanism must also allocate its privacy budget among different code fragments. This results in a trade-off where fragments with a larger share of the privacy budget use smaller amounts of noise.  Picking the optimal (in terms of utility) way of adding randomness and allocating the privacy budget is also a non-trivial task (this is especially true for SVT~\cite{ninghuisparse}).
\end{itemize}

Most current tools focus on checking the privacy properties of algorithms. For example, verification tools have been developed to mechanically (and sometimes, automatically) prove that a (correct) privacy mechanism satisfies differential privacy~\cite{lightdp, shadowdp, Aws:synthesis, Barthe12,EasyCrypt,BartheICALP2013,Barthe16,BartheCCS16}. Counterexample detectors for differential privacy~\cite{Ding2018CCS,Bichsel2018CCS,diffproptest,dpsniper} can find evidence that an (incorrect) privacy mechanism fails to satisfy its claimed privacy levels. Moreover, a few tools can  combine both functionalities: either proving a mechanism is correct or finding a counterexample~\cite{checkdp,barthe2020, FarinaCG2021}. While these tools are invaluable for ensuring correctness of privacy mechanisms, they all require a putative differentially private algorithm as a starting point.

Recently, Roy et al.~\cite{learning} took a step further: they proposed a tool called KOLAHAL for automatically learning an accurate and differentially private mechanism \emph{given a mechanism sketch provided by a domain expert}. In other words, their approach synthesizes how much noise should be added in pre-specified locations. Note that it does not determine \emph{where} the noise should be added. It also cannot synthesize mechanisms that use their privacy budget adaptively. An example of such a mechanism is a recently proposed variant of SVT, called Adaptive Sparse Vector with Gap~\cite{freegap}. This mechanism has extra flexibility for saving privacy budget on some queries, allowing it to keep iterating until its privacy budget is exhausted.

In this paper, we present \tool, the first \emph{fully automated} approach that can synthesize an accurate and differentially private program from a given non-private (noiseless) program. Significantly, \tool employs a novel inference algorithm to automatically generate a mechanism sketch from a non-private program. We formalize the synthesis problem as a constrained optimization problem: maximizing \emph{utility} while simultaneously satisfying \emph{privacy} constraints in a transformed version of the mechanism sketch. \tool then uses a counterexample-guided synthesis (CEGIS) loop~\cite{CEGIS} and an optimizer like Particle Swarm Optimization (PSO)~\cite{pso} to synthesize and  optimize the mechanism. Compared with KOLAHAL~\cite{learning}, the new optimization approach is shown to be more efficient. In some cases, KOLAHAL can take 900 to 5460 seconds to synthesize a mechanism, while \tool can successfully synthesize an equivalent or more accurate version in 10 to 120 seconds.

Moreover, \tool is equipped with a novel feature called a \emph{while-private} loop, written as $\whilepriv{e}{c}$. Semantically, the while-private loop (after synthesis) executes $c$ whenever $e$ evaluates to $\true$ and \emph{as long as the dynamically tracked privacy budget has not been depleted}. Notably, this feature allows \tool to synthesize sophisticated mechanisms such as Adaptive Sparse Vector with Gap~\cite{freegap} that try to minimize the amount of privacy budget spent in each loop iteration, and hence keep iterating until the privacy budget has been depleted. %
To the best of our knowledge, \tool is the first program synthesizer that can automatically generate such sophisticated mechanisms.\footnote{Note that \emph{while-private} is a programmer hint that the while loop should be executed in a best-effort way (a hallmark of the sparse vector family of privacy mechanisms) rather than exactly as many times as the non-private version would execute.}

We evaluated \tool on standard benchmarks that consist of various privacy mechanisms. For each privacy mechanism, we removed the randomness in it and asked \tool to automatically synthesize a differentially private version. In all cases, \tool was able to synthesize an equivalent or even more accurate version compared with the baseline. For adaptive mechanism that uses while-private loop, program synthesis is more complicated. But \tool was still able to synthesize private and accurate mechanisms.

In summary, this paper makes the following contributions:
\begin{enumerate}
\item  \tool, the first fully automated tool that can synthesize an accurate and differentially private mechanism from a noiseless non-private program.

\item A novel inference algorithm that automatically generates a mechanism sketch (i.e., code with noise of unknown scales added to automatically selected program locations) from a non-private program (Section~\ref{sec:sketch}).

\item A customized CEGIS loop that incrementally optimizes the tentative mechanism while generating its privacy proof (Section~\ref{sec:generation}).

\item A novel \emph{while-private} feature that allows \tool to synthesize adaptive privacy mechanisms (Section~\ref{sec:utility_by_example}).

\item Case studies and experimental comparisons between \tool and KOLAHAL~\cite{learning}. In addition to being able to synthesize more programs,  \tool also shows improvements on mechanims that both approaches can synthesize. In the benchamrks, \tool generated identical or more accurate mechanisms within a considerably shorter amount of time (Section~\ref{sec:evaluation}). %

\end{enumerate}

%% file: sec_preliminaries.tex
\subsection{Differential Privacy}
In this paper, 
we focus on \emph{pure} differential privacy~\cite{dwork06Calibrating}.
Intuitively, a data analysis $A$ satisfies differential privacy if and only if for any dataset $D$, adding, removing, or changing a record in $D$ has little impact on the analysis result. Therefore, a differentially private analysis reveals little about any data record being analyzed. Each analysis is built out of atomic components called differentially private mechanisms (privacy mechanisms for short). These components themselves satisfy differential privacy.\footnote{In general, the privacy parameter of the analysis is upper bounded by the sum of the individual privacy parameters of the mechanisms \cite{pinq}.}

More formally, we say that two datasets $D, D^\prime\in \algoinput$ are \emph{adjacent}, written $D\sim D'$, when they only differ on one record.
To offer privacy, a differentially private mechanism (or analysis), say $\mechanism: \algoinput \rightarrow \algooutput$, injects carefully calibrated random noise during its computation. We call the execution of $\mechanism$ on $D$, written $\mechanism(D)$, the \emph{original execution} and its execution on (neighboring) dataset $D^\prime$, written $\mechanism(D^\prime)$, the \emph{related execution}. Intuitively, $\mechanism$ (or $A$) is $\epsilon$-differentially private for some constant $\priv$ if for any possible output $o\in \algooutput$, the ratio between the probabilities of producing $o$ on $D$ and $D'$ is bounded by $e^\priv$:
\begin{definition}[Pure Differential Privacy \cite{Dwork06diffpriv}]
    \label{def:diffpriv}
    Let $\epsilon \geq 0$. A probabilistic computation $\mechanism: \algoinput \rightarrow \algooutput$ is $\priv$-differentially private  if %
      $\forall D\sim D^\prime$ (where $D,D^\prime \in \algoinput$) and $\forall o\in \algooutput$, we have $$\prob[M (D) = o]\leq e^\priv \prob[M (D') = o]$$ %
\end{definition}  

A differentially private analysis $A$ interacts with a dataset through  one or multiple privacy mechanisms that take a list of queries and their exact answers as input, and produce a differentially private (noisy) aggregation of them. An important factor to determine the amount of noise needed for privacy is the \emph{sensitivity} of queries, which intuitively quantifies the maximum difference of the query results on adjacent databases. We use a vector $(q_1,q_2,\ldots)$ to denote the exact query answers from running a sequence of queries on a dataset and say that each query answer $q_i$ has a \emph{sensitivity} of $\Delta_i$ if its corresponding query has a \emph{global sensitivity} of $\Delta_i$:
\begin{definition}[Global Sensitivity \cite{diffpbook}]
    \label{def:sensitivity}
    The global sensitivity $\Delta_f$ of a query $f$ is $\sup_{D\sim D'} \abs{f(D)-f(D')}$.
\end{definition} 
Similar to dataset adjacency, we say two vectors of query answers are \emph{adjacent}, written $(q_1,q_2,\ldots)\sim (q_1',q_2',\ldots)$, when $\forall i.~|q_i-q_i'|\leq \Delta_i$. Moreover, a privacy mechanism $\mechanism$ satisfies $\epsilon$-differential privacy if for all pairs of adjacent query answers $(q_1,q_2,\ldots)\sim (q_1',q_2',\ldots)$ and all outputs $o\in \algooutput$, we have $\prob[\mechanism(q_1,q_2,\ldots,\text{params}) = o] \leq e^\priv \prob[\mechanism(q_1',q_2',\ldots,\text{params}) = o]$, where params represent data-independent parameters (e.g., the value of $\priv$) to $\mechanism$.
As the goal of this paper is to synthesize privacy mechanisms, we assume that the sensitivity of inputs are either manually specified or computed by sensitivity analysis tools (e.g.,~\cite{Fuzz,DFuzz}).

One popular privacy mechanism is the Laplace Mechanism~\cite{dwork06Calibrating}, which adds Laplace noise to query answers.
\begin{theorem}[Laplace Mechanism \cite{dwork06Calibrating}]
Let $\lapm(n)$ be a sample from the Laplace distribution with mean 0 and scale $n$. The Laplace Mechanism  takes as input a query answer $q$ with sensitivity $\Delta_q$,  and a privacy parameter $\epsilon$. It outputs $q + \lapm{(\Delta_{q}/\epsilon)}$ and it satisfies $\epsilon$-differential privacy. 
\end{theorem}

In this paper, we will use Laplace noise to also synthesize more sophisticated privacy mechanisms.

\subsection{Randomness Alignment}
\label{sec:alignment}
To synthesize a privacy mechanism, we need to reason about its correctness (i.e., it must satisfy pure differential privacy with a given privacy parameter $\epsilon$). To mechanize the correctness reasoning, we adopt the \emph{Randomness Alignment} technique, a simple yet powerful proof technique that enables various verification tools and counterexample detectors~\cite{lightdp,shadowdp,checkdp}.

Consider a privacy mechanism $\mechanism$ and an arbitrary pair of adjacent vectors of query answers $(q_1,q_2,\ldots)\sim (q_1',q_2',\ldots)$. A randomness alignment is a function $\phi: \mathbb{R}^{\infty}\rightarrow \mathbb{R}^{\infty}$ that maps random samples used by an execution of $\mechanism$ on $(q_1,q_2,\ldots)$ to random samples used by the adjacent execution of $\mechanism$ on $(q_1',q_2',\ldots)$ such that both executions produce the same output.

For example, consider the mechanism $\mechanism(x)=x+\lapm(\priv)$ that adds Laplace noise to a query answer $x$ of sensitivity 1. Then, given any pair of adjacent query answers $q\sim q'$, the function $\phi(r)=r+q-q'$ is an alignment. The reason is that for any possible Laplace random sample $\eta$ generated by $M(q)$, we have $q+\eta=q'+(\eta+q-q')=q'+\phi(\eta)$ (i.e., $M(q')$ produces the same result when its Laplace sample is $\phi(\eta)$). 

To finish the privacy proof, we note that for Laplace distribution $\lapm(\priv)$, the ratio of the probabilities of sampling $\eta$ and $\phi(\eta)$ is bounded by $e^{\priv \max_{r\in \mathbb{R}} |\phi(r)-r|}=e^{\priv \max_{r\in \mathbb{R}} |q-q'|}\leq e^\priv$. Hence, the \emph{privacy cost}, the natural log of this ratio, is bounded by $\epsilon$.

In general, it is useful to treat the privacy cost as a function of the alignment needed for each sampling instruction. For each sampling instruction $\eta=\lapm(r)$, we define the \emph{distance} of $\eta$, written as \distance{$\eta$}, as $\phi(\eta)-\eta$\footnote{Here we abuse notation slightly by applying $\phi$ point-wise, letting $\phi(\eta)$ be the random sample $M$ should use in place of $\eta$ in the adjacent execution.}. Then, the privacy cost of aligning the sample $\eta$ is bounded by $\frac{|\distance{$\eta$}|}{r}$. To find the overall privacy cost (i.e., the $\epsilon$ in pure differential privacy), we then take the summation of privacy cost of each sample generated in program execution, due to the Composition Theorem of pure differential privacy~\cite{diffpbook}. We note that since we can align each sample individually, randomness alignment is also applicable to sophisticated mechanisms where the composition theorem falls short~\cite{lightdp,shadowdp}. This is a key to automated synthesis of a variety of mechanisms studied in this paper.

\subsection{Particle Swarm Optimization (PSO)}
Prior tools using the Randomness Alignment technique (e.g.,~\cite{lightdp, shadowdp, checkdp}) focus on \emph{privacy} only; they model privacy proof as a constraint-solving problem which is solved by an external SMT solver. 
However, %
synthesizing DP mechanism is better described as a \emph{constrained optimization} problem: maximizing \emph{utility} among  various candidates that have the same overall differential privacy parameter $\epsilon$. 

In this paper, we use Particle Swarm Optimization (PSO)~\cite{pso} to help with the synthesis. PSO is a meta-heuristic optimization algorithm that is inspired by swarm behaviors such as birds in nature. It deploys a large population of candidate solutions (``particles'') in the search space and the particles move around iteratively to find the best location. For each iteration, each particle updates its position and velocity according to a mathematical formula consisting of its own local best position, the swarms' best position and its previous velocity. By adopting this strategy, the entire swarm is guided towards the best solutions. PSO makes no assumption about the problem being optimized and is suitable for very large search spaces. This is well suited for our complex, non-differentiable optimization problem, which makes other gradient-based optimization methods inapplicable. Specifically for the synthesis task, each candidate mechanism in the search space corresponds to a particle in PSO, and the instantiations of the sketch mechanism serves as its position. For each iteration, every candidate explores the search space by changing itself slightly according to the current global best candidate (with the best utility), its own local best in history and the amount of changes from previous iterations. The global best solution is returned after a number of iterations.

\subsection{Sparse Vector Technique (SVT)}\label{sec:svt}
\input{figures/svt}

In this paper, we use Sparse Vector Technique (SVT)~\cite{diffpbook} and its variants as running examples. Given a sequence of queries, SVT tries to find the first $N$ queries whose query answers are likely\footnote{The uncertainty is introduced by privacy requirements.} to be above a publicly known threshold $T$. When privacy is not a concern, the pseudo code of SVT's basic functionality is shown in Figure~\ref{fig:svt} (we call it \code{SVTBase}). For now, we can safely ignore the function signature. \code{SVTBase} checks each exact query answer: it outputs $\true$ (resp. $\false$) if the query answer is above (resp. below) the threshold until $N$ $\true$ outputs are produced.

To enforce differential privacy, SVT adds \emph{carefully calibrated} independent Laplace noise both to the threshold ($T$) and each query answer ($q[i]$). The pseudo code is shown in Figure~\ref{fig:svt} (we call it \code{SVT}), where the changes are highlighted. The sampling instruction $\lapm(r)$ draws one sample from the Laplace distribution with mean 0 and scale factor of $r\in \mathbb{R}$. For each query, the mechanism outputs $\true$ if the \emph{noisy} query answer is above the \emph{noisy} threshold; otherwise it outputs $\false$. It is well-known that SVT satisfies $\epsilon$ differential privacy~\cite{diffpbook}.

%% file: figures/svt.tex
\begin{figure}
\raggedright
\noindent\rule{\linewidth}{0.8pt}
\funsigfour{SVTBase}
{\code{T,N,size}: \tyreal, \code{q}: \tylist~$\synth{\tyreal}$}
{(\code{out}: \tylist~\tyreal), \bound{\priv}}
{\alldiffer $\land$ \code{N < \code{size / 5}}}
\algrule
\begin{lstlisting}[frame=none,escapechar=@]
i := 0; count := 0;
while (i < size $\land$ count < N)@\label{line:svt:while}@
  if ($\text{q[i]} \geq T$) then @\label{line:svt_branch}@
    out := true::out;
    count := count + 1;
  else
    out := false::out;
  i := i + 1;
\end{lstlisting}
\noindent\rule{\linewidth}{0.8pt}
\funsigthree{SVT}
{\code{T,N,size}: \tyreal, \code{q}: \tylist~\tyreal}
{(\code{out}: \tylist~\tyreal)}
\algrule
\begin{lstlisting}[frame=none,escapechar=@]
i := 0; count := 0;
@\annotation{\eta_1 := \lapm{(3/\priv)}}@
@\annotation{\tT := T + \eta_1;}@
while (i < size $\land$ count < N)
  @\annotation{\eta_2 := \lapm{(3N/\priv)};} \label{line:gapsvt_eta2}@
  if ($\text{q[i]}$ @\annotation{+ \eta_2}@ $\geq$ @\annotation{\tT}@) then
    out := true::out;
    count := count + 1;
  else
    out := false::out;
  i := i + 1;
\end{lstlisting}
\noindent\rule{\linewidth}{0.8pt}
\caption{Sparse Vector Technique. \label{fig:svt}}
\end{figure}

%% file: sec_overview.tex
\subsection{Challenges}
\label{sec:challenges}
The goal of this paper is to \emph{automatically synthesize} a differentially private program (e.g., function $\code{SVT}$) from a base program that is not necessarily differentially private (e.g., function $\code{SVTBase}$). Like other program synthesis techniques~\cite{CEGIS,gulwani2011}, the synthesized program must implement similar functionality to the original program / specification. Since a privacy mechanism injects noise to offer privacy, this can be more precisely stated as: any output of the original program is still possible for the synthesized program.

What makes \tool distinguished from other program synthesizers is its capability of synthesizing a \emph{private} and \emph{useful} counterpart of the original program:
\begin{itemize}
    \item Privacy: the synthesized program needs to inject sufficient noise in the right places to satisfy pure differential privacy, as formally defined in Definition~\ref{def:diffpriv}.

    \item Utility: the synthesized program needs to carefully calibrate the injected noise to make the randomized outputs useful (i.e., to make the outputs ``close'' to the ones from the original program). This involves choosing the correct noise scales (including using no noise wherever it is safe to do so).
\end{itemize}

Next, we highlight the main challenges in both aspects.
\paragraph{Privacy} Developing differentially private mechanisms is a nontrivial task: injecting sufficient amount of noise in the right places and then proving correctness is notoriously tricky. For instance, Lyu et al.~\cite{ninghuisparse} catalog several incorrect variants of SVT, where each variant slightly modifies the functionality and/or injected noise of function \code{SVT} in Figure~\ref{fig:svt} (for now, safely ignore the annotations in the function signature). While the changes are minimal, the incorrect variants fail to meet their claimed differential privacy guarantees. For example, one variant tweaks the mechanism to output the noisy query answer when it is above the threshold. That is, it changes 
Line 7 of \code{SVT}
by replacing $\true$ with  $q[i]+\eta_2$. %
As a result, it fails to satisfy $\epsilon$-differential privacy for any value of $\epsilon$~\cite{ninghuisparse}.

\paragraph{Utility}

What makes synthesizing differentially private mechanisms even more challenging is that we also need to add as little noise as possible while maintaining the desired privacy levels (otherwise the noisy outputs may not be useful). For example, in the simplest case, if we increase the scale of noise injected at Lines 2 and 5 in SVT (Figure~\ref{fig:svt}), the mechanism is still $\epsilon$-differentially private. However, the extra randomness reduces the accuracy of SVT. 
\input{figures/svt_alt}
Furthermore, utility is also affected by \emph{where} the noise is added. %
For example, an alternative way of making function $\code{SVTBase}$ $\priv$-private is shown in Figure~\ref{fig:svtalt}. Compared with SVT, SVT-ALT does not add any noise to the threshold $T$; instead, it injects Laplace noise $\lapm{(size/\priv)}$ (rather than $\lapm{(3N/\priv)}$) to each query answer. This provides the same privacy guarantees (SVT and SVT-ALT both satisfy $\epsilon$-differential privacy for the same value of $\epsilon$). However, since $N$ is typically much smaller than $\texttt{size}$ (the total number of queries), SVT-ALT injects significantly more noise into its computation.

Handling these kinds of decisions during the synthesis process is a highly non-trivial task and requires deep understanding of the privacy cost introduced by each sampling instruction. For example,
SVT and its correct variants~\cite{diffpbook,ninghuisparse,ashwinsparse,freegap} have the interesting property that outputting $\false$ \emph{does not} incur any privacy cost (i.e., the costs\footnote{The privacy cost of the threshold is $\epsilon/3$ and each of the $N$ $\true$ outputs incurs a privacy cost of $2\epsilon/(3N)$.} are only incurred for making the threshold noisy and for outputting $\true$). On the other hand SVT-ALT is too naive and incurs a privacy cost of $\epsilon/\texttt{size}$ for each iteration of the while loop (for a total cost of $\epsilon$). %
\begin{figure*}[t]
  \includegraphics[width=.8\textwidth]{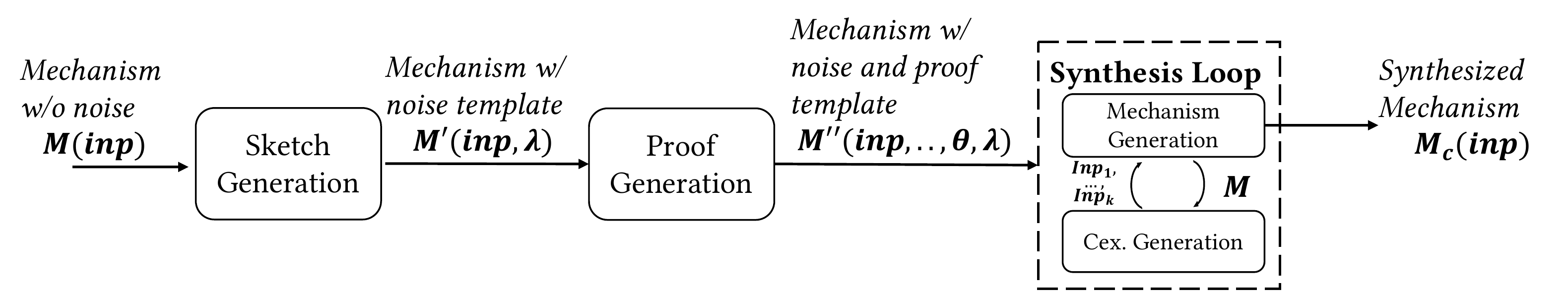}
  \vspace{-8pt}
  \caption{Overview of \tool.}\label{fig:overview}
\end{figure*}

Finally, in many mechanisms (including SVT) and its variants, one needs to decide how to divide up a total privacy budget $\priv$ among different parts of the mechanism (i.e., what should the privacy cost of each part of the mechanism be). In the case of SVT, a synthesizer would decide how much of the budget should be consumed by adding noise to threshold $T$ and how much should be consumed by the while loop. This is equivalent to deciding how much noise should be used for the threshold and how much should be used for the noisy query answers. In Figure~\ref{fig:svt}, the noise scale for the threshold is $\sigma_1=3/\epsilon$ while the noise scale for each query answer is $\sigma_2=3N/\epsilon$. However, any choice of $\sigma_1,\sigma_2$ that satisfies $1/\sigma_1 + 2N /\sigma_2=\epsilon$ will result in $\epsilon$-differential privacy \cite{ninghuisparse}.
As shown by Lyu et al.~\cite{ninghuisparse}, an approximately optimal ratio of $\sigma_1:\sigma_2$ is $1:(2N)^{2/3}$. %

\subsection{Approach Overview}

To synthesize a privacy mechanism, \tool adds proper amount of noise to the original program. This naturally involves two tasks: (1) finding program locations to add random noise to, and (2) finding the amount (scale) of each noise. 
Accordingly, \tool synthesizes a privacy mechanism as 
shown in Figure~\ref{fig:overview}.

\paragraph{Phase 1: Sketch Generation (Section~\ref{sec:sketch})}
In Phase 1, \tool generates a \emph{sketch mechanism} with candidate locations for noise. The sketch mechanism might contain more locations for noise than needed, as the unnecessary ones will eventually be optimized away in Phase 2. Moreover, each noise location $\eta_i$ is paired with a scale template $\scale_i$ which consists of a set of unknown scale holes $\scalehole$ to be synthesized in Phase 2. We use $M'(inp,\scalehole)$ to denote such a sketch mechanism with unknown scale holes.

\paragraph{Phase 2: Synthesis Loop (Section~\ref{sec:synthesis})}
Due to the tension between privacy and utility, mechanism synthesis cannot proceed without privacy in mind. Hence, \tool next generates a transformed relational program with both scale templates $\scale$ containing holes $\scalehole$, and proof templates (in the form of alignments) $\alignment$ containing holes $\hole$ to be synthesized. 
Next, \tool employs a customized CEGIS loop that iteratively refines a candidate mechanism (i.e., an instantiation of $\hole$ and $\scalehole$) by generating more and more counterexamples (i.e., inputs that violates privacy constraints). %

The CEGIS loop consists of two components. The counterexample generation component starts with a null mechanism (with $\hole = \vec{0}$ and $\scalehole = \vec{1}$) and first searches for a counterexample (i.e., inputs) that \emph{maximizes} the total number of privacy violations. The reason behind the optimization goal is the following: CEGIS benefits greatly from a good set of counterexamples; intuitively, a counterexample that violates maximum number of privacy constraints serves as better guides than others. 

With a set of counterexamples, the mechanism generation component searches for a mechanism (i.e., an instantiation of the mechanism template) that \emph{maximizes} utility while still being private. More specifically, the utility is defined both for privacy and accuracy:
\begin{itemize}
    \item \textbf{Privacy.} A mechanism must be private for all previously seen counterexamples. Hence, any mechanism that is deemed as non-private on counterexamples has a negative utility score.
    
    \item \textbf{Accuracy.} \tool is parameterized by either a default utility function (sum of variances), or a user-provided one. The utility function  is used as the quality metric of each private candidate.
\end{itemize}

Once \tool finds a mechanism where no counterexamples can be found, the CEGIS loop terminates and \tool sends the mechanism to a verifier (we use CPAChecker~\cite{beyer2011cpachecker}). Note that although we did not encounter any incorrect synthesized mechanism in our experiments, verification is needed in general as an optimizer might miss a solution when one exists.

%% file: figures/svt_alt.tex
\begin{figure}
\raggedright
\noindent\rule{\linewidth}{0.8pt}
\funsigthree{SVT-ALT}
{\code{T,N,size}: \tyreal, \code{q}: \tylist~\tyreal}
{(\code{out}: \tylist~\tyreal)}
\algrule
\begin{lstlisting}[frame=none,escapechar=@]
i := 0; count := 0;
while (i < size $\land$ count < N)
  @\annotation{\eta_2 := \lapm{(size/\priv)};}@
  if ($\text{q[i]}$ @\annotation{+ \eta_2}@ $\geq T$) then
    out := true::out;
    count := count + 1;@\label{line:svt:count}@
  else
    out := false::out;@\label{line:svt:false}@
  i := i + 1;
\end{lstlisting}
\noindent\rule{\linewidth}{0.8pt}
\caption{An alternative way of making \code{SVTBase} $\epsilon$-private.} \label{fig:svtalt}
\end{figure}

%% file: sec_sketch.tex
As discussed in Section~\ref{sec:overview}, \tool synthesizes a DP mechanism in two phases. In this section, we first show the syntax of its source and target languages. Then, we propose novel algorithms to identify potential violations of privacy in the source code, and then, to inject noise at proper locations to form a program sketch to be further analyzed in Phase 2 (Section~\ref{sec:synthesis}).

\subsection{Syntax of Source and Target Program}
\input{figures/syntax.tex}

\paragraph{Source Language}
The syntax of \tool source code is listed in Figure~\ref{fig:syntax}. The source language models an expressive imperative language with the following standard features:
\begin{itemize}
\item Values of real numbers, Booleans and operations on them;

\item Ternary expressions $\ternary{\bexpr}{\nexpr_1}{\nexpr_2}$, which returns $\nexpr_1$ (resp. $\nexpr_2$) when $\bexpr$ evaluates
to $\true$ (resp. $\false$);

\item List of values as well as append (::) and projection ([]) operations on lists. Note that all lists are initialized to be empty.

\item No-op commands ($\skipcmd$), assignments, sequential commands ($c_1;c_2$), return commands, if branches and while loops. 
\end{itemize}

One novel feature of the source language is a while-private loop written as $\whilepriv{e}{c}$; it requests the synthesizer to synthesize an \emph{adaptive} privacy mechanism (e.g., Adaptive Sparse Vector with Gap~\cite{freegap}) that runs $\whilecmd{e}{c}$ until the privacy budget is exhausted. This powerful feature allows the synthesized privacy mechanism to adaptively control the number of outputs based on the remaining privacy budget, in order to increase the amount of queries that they can process. We show how to synthesize the Adaptive Sparse Vector with Gap mechanism in Section~\ref{sec:utility_by_example}.

Finally, the source language requires a few user-provided privacy specifications that the synthesizer should obey, including private inputs and their sensitivity\footnote{Determining the sensitivity of queries is crucial to produce an appropriate noise scale. Here, we assume that this information is provided by the user, as the sensitivities of simple queries, such as sum, mean and median, are fairly easy to compute as demonstrated in \cite{diffpbook}. For more complex queries, users can either derive manually or use sensitivity analysis tools (e.g., \cite{rappor,ashwin08:map}) to calculate sensitivity.}, the desired privacy bound (i.e., $\epsilon$ in $\epsilon$-differential privacy), as well as assumptions on the query answers. While we do not formalize the syntax of such specification, we use $\synth{\code{type}}$ to denote
private input of some $\code{type}$, $\bound{\priv}$ to denote the privacy budget, and specify sensitivity on private inputs ($\distance{$x$}$ represents the sensitivity of $x$) and other assumptions on inputs as program precondition. For example, the source program $\code{SVTBase}$ in Figure~\ref{fig:svt} specifies that query answers $q$ are the only private inputs and their sensitivity is 1. Moreover, the mechanism assumes that $N$ is much smaller than $size$, and the goal is to synthesize an $\epsilon$-differentially private mechanism.

\paragraph{Target Language}
The goal of \tool is to synthesize a randomized mechanism that both preserves the source program's semantics and offers $\epsilon$-differential privacy (where $\epsilon$ is annotated in the source program). Hence, the target language (shown in Figure~\ref{fig:syntax}) is similar to the source language, with a few important changes:
\begin{itemize}
    \item The target language is probabilistic: it extends the (deterministic) source language with random variables $\eta$ and sampling commands, written as $\eta:=\lapm(\nexpr)$.
    
    \item The target language excludes the (non-executable) while-private loops; such loops in the source code are replaced by fully synthesized standard loops that terminate the loop whenever the privacy budget is exhausted.
\end{itemize}

Consider Figure~\ref{fig:svt}. Function $\code{SVT}$ is the target program synthesized from the source program $\code{SVT-Base}$. Note that they are very similar, but function $\code{SVT}$ properly injects noise at various locations to satisfy $\epsilon$-differential privacy.

\subsection{Adding Noise Locations to Source Code}

The first step of \tool is to find a \emph{set of program locations} in the source program where extra noise is needed. In this step, the primary concern is \emph{privacy}; in other words, the lack of randomness in the source program violates differential privacy. Hence, we use static program analysis to (1) identify \emph{where} privacy is violated in the source code, (2) infer a set of variables that might require randomness, and (3) instrument the source code to inject noise to the identified variables.

\subsubsection{Identify Violations of Differential Privacy}
Recall that \tool is built on the Randomness Alignment technique (Section~\ref{sec:alignment}) to reason about privacy. Hence, instead of analyzing properties on distributions directly, as stated in Definition~\ref{def:diffpriv}, we over-approximate ``Violations of Differential Privacy'' as ``Violations of Alignment Requirements''. Recall that randomness alignment requires that when running on a pair of adjacent private inputs, a program will produce identical outputs. Since the source code has no randomness, this requirement can be formalized as the standard non-interference property~\cite{noninterference}. Hence, we use a static taint analysis (e.g.,~\cite{sm-jsac,vsi96,Hunt:flowsensitive}) to identify violations in the source code:
\begin{itemize}
    \item Initially, only the private inputs are tainted.
    
    \item The analysis tracks all \emph{explicit flows} in the program.
    
    \item The analysis does not track, but reports all \emph{implicit flows}, where a tainted value is used in a branch condition.
    
    \item The analysis reports all outputs with a tainted value.
\end{itemize}

For example, since query answers $q$ are the only tainted inputs in  \code{SVTBase} (Figure~\ref{fig:svt}), the taint analysis finds one violation of privacy at Line 3, where the branch condition uses a tainted value $q[i]$. Since the taint analysis is standard, we omit the details here.

\subsubsection{Identify Offending Variables}
The static taint analysis returns a set of offending assignments $x:=e$ and offending branches $\ifcmd{e}{c_1}{c_2}$, where $e$ is tainted. We use $\mathbb{E}$ to represent the set of expressions that are either on the RHS of offending assignments, or in the branch condition of offending branches. Next, we need to infer a set of variables, that when randomized, will allow randomness alignment to exist on the randomized code. We call such a set of variables \emph{offending variables}.

Consider the offending branch in our running example:
\begin{lstlisting}[numbers=none, frame=none, escapechar=@]
@\ifcmd{q[i] $\geq$ T}{...}{...}@
\end{lstlisting}
where $q[i]$ is tainted while T is not. To make the branch outcome identical on two adjacent inputs $q[i]\sim q'[i]$, we can either inject noise to $q[i]$, or to $T$, or to both. While all options can allow the offending branch to be aligned, the difference will show up when we analyze their corresponding utility. For example, adding noise to $T$ is crucial to make SVT useful; intuitively, it allows the noisy $T$ to be reused across different loop iterations, which results in a less noisy program. We defer the discussion on utility to Section~\ref{sec:util}.

Based on the insight above, we define all variables used in any $e\in \mathbb{E}$ as offending variables. Note that by definition, the set of tainted variables is always a subset of offending variables.

\subsubsection{Instrument Source Code with Extra Noise}

\input{figures/svt_sketch.tex}

Finally, \tool injects noise with \emph{unknown scales} (to be synthesized in later stages) to the source code. In particular, it injects Laplacian noise both at the definition of an offending variable, as well as right before its corresponding uses in an offending command. While adding noise to both locations might seem  unnecessary at this point, \tool eventually uses a utility optimizer (Section~\ref{sec:util}) to remove unnecessary noise in the code sketch.

Moreover, as the scale of each Laplacian noise is unknown at this point, we replace them with \emph{scale templates} as follows:
\[
(\scalehole_0 + \sum_{v_i \in \varset} \scalehole_i \times v_i) / \epsilon \text{ with fresh } \scalehole_i
\]
where $\varset$ contains all \emph{non-private} function parameters (as making scale private could violate privacy directly by revealing distribution statistics).
Return to our running example of SVT, the code sketch with extra noise is shown in Figure~\ref{fig:svt_sketch} where all changes are highlighted. Notably, the sketched function explicitly adds scale parameters $\lambda$ (we use $\lambda_i$ instead of $\lambda[i]$ for better readability) as extra inputs to be optimized later. No noise is injected at Line 2 for $q[i]$, essentially an iterator of $q$, as it is not in scope at that point. 

Hereafter, we use $M(inp)$ and $M'(inp, \lambda)$ to represent the original program with inputs $inp$ and mechanism sketch with scale parameters $\lambda$ respectively.

\section{Synthesis and Optimization}
\label{sec:synthesis}

In Phase 2, \tool completes program synthesis with two sub-goals: 
\begin{itemize}
    \item It synthesizes and optimizes the randomness alignment of each sampling instruction; a sampling instruction with alignment $0$ implies that the instruction can be removed without violating differential privacy.
    
    \item It synthesizes and optimizes the scales $\lambda$ in the sketch code from Phase 1 to offer good utility.
\end{itemize} 

The main challenge is that instead of synthesizing \emph{some} privacy proof (as done in prior work with proof synthesis~\cite{checkdp,Aws:synthesis}) or \emph{optimize} scales with given randomness locations (as done in~\cite{learning}), our goal is to synthesize and optimize both the proof (with fewest randomness locations) and scales. 

We first introduce the optimization problem without any while-private loop in source code and assume a default utility function that minimizes sum of variances. Then, we propose a synthesis loop to optimize alignments and scales simultaneously. Finally, we generalize the approach to optimize sketch code with while-private loops and customized utility functions.  

\subsection{Mechanism Synthesis Problem}
\label{sec:modeling}
\input{sec_modeling}

\subsection{Mechanism Optimization Problem}\label{sec:generation}
Recall that the goal of \tool is to generate an \emph{accurate} and \emph{private} mechanism. That is, for a search space of alignment holes $\Theta$ and scale holes $\Lambda$, the constrained optimization problem is defined follows:
\begin{equation}
  \begin{aligned}
    \max_{(\hole,\scalehole) \in \Theta \times \Lambda} & \quad \code{Utility}(M', \hole, \scalehole) \\
  \textrm{s.t.} & \quad  \forall inp. \text{all assertions in } M'' \text{ pass }   \\\nonumber
  \end{aligned}
\end{equation}

\begin{figure}
  \includegraphics[width=\linewidth]{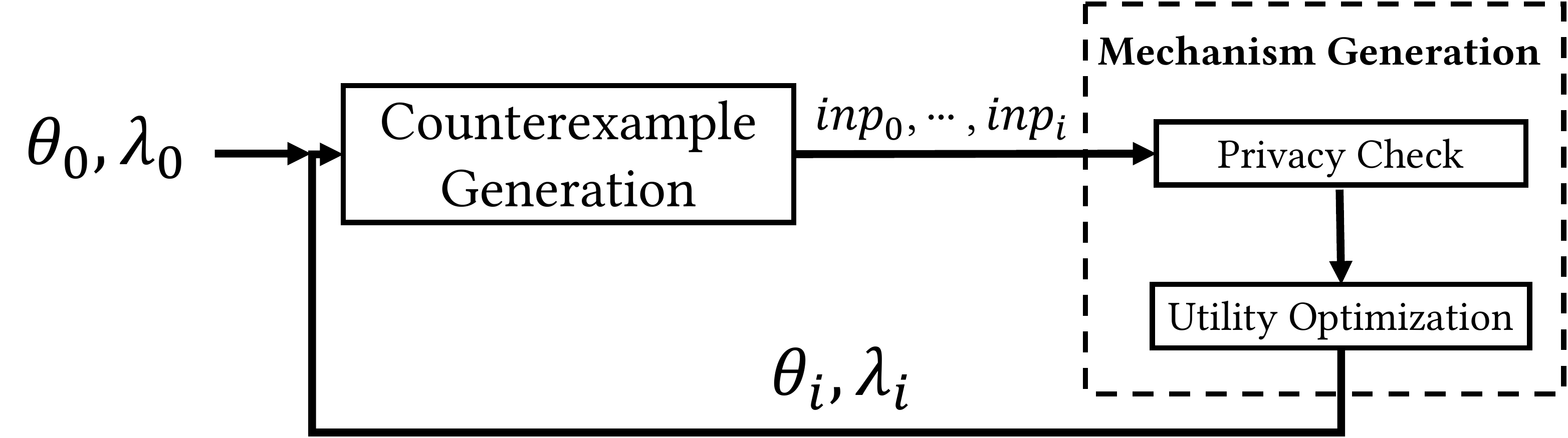}
  \caption{Overview of the search loop.}\label{fig:loop}
\end{figure}

To find alignment holes ($\hole$) and scale holes ($\lambda$) according to the optimization problem above, \tool uses a customized Counterexample-Guided Inductive Synthesis (CEGIS)~\cite{CEGIS} loop, as illustrated in Figure~\ref{fig:loop}. Each synthesis iteration contains two steps:
\begin{itemize}
    \item With a candidate mechanism (initialized with null mechanism of $\hole_0 = \vec{0}, \scalehole_0 = \vec{1}$), the ``counterexample generation'' component tries to find inputs $inp$ that ``break'' the privacy requirements (i.e., assertion violations in $M''$).
    
    \item With a set of counterexamples seen so far, the ``mechanism generation'' component synthesizes a privacy mechanism by optimizing the utility objective function (we use PSO as a black-box optimization technique in this paper) while satisfying all previously-generated counterexamples.
\end{itemize}
The CEGIS loop terminates when no counterexamples can be generated; then, the final privacy mechanism is returned.

Compared with the ``bi-directional'' search loop of CheckDP~\cite{checkdp} that improves both privacy proof and counterexamples simultaneously, the CEGIS loop in Figure~\ref{fig:loop} is more standard, as
there is no need to improve counterexamples for \tool. Hence, the use of ``bi-directional'' CEGIS loop is not necessary.

\paragraph{Discussion on Soundness}
Note that since most optimizers (including PSO~\cite{pso} that \tool uses) are unsound (i.e., they might miss a solution when one exists),  the synthesized privacy program might be (in rare cases) non-private. To ensure soundness, the synthesized mechanism can be further verified by sound tools like CheckDP~\cite{checkdp}. If verification fails, the counterexamples generated from CheckDP can be passed back to the CEGIS loop to continue the search. 
In practice, we did not experience any such unsound cases by running separate verification passes in CheckDP; we leave the integration of \tool and CheckDP as future work.

\subsubsection{Counterexample Generation}
Given a candidate mechanism instantiated with some $\hole,\lambda$, as well as a transformed mechanism with explicit alignments $M''(inp,\distance{$inp$},sample,\hole,\scalehole)$, a counterexample $C$ is defined as a solution of the following term:
\[
\exists inp,\distance{$inp$},sample.~\text{ some assertions in } M''(inp,\distance{$inp$},sample,\hole,\scalehole) \text{ fail}.
\]

We note that this naive definition treats all counterexamples equally: two distinct counterexamples which violate 1 and 100 assertions respectively are both acceptable. To quantify and optimize the qualities of counterexamples (for better performance), we slightly modify the mechanism $M''$ to return the total number of assertion violations and use an optimizer to find a counterexample according to the following metric:
\[
  \max_{inp,\distance{$inp$},sample} M''(inp, \distance{$inp$}, sample, \hole, \scalehole)
\]

Consider the transformed program of our running example in Figure~\ref{fig:svt_trans} with a null mechanism ($\hole=\vec{0}, \scalehole=\vec{1}$) for bootstrapping the process. The optimizer tries to find a counterexample that fails as many assertions as possible. Since no alignments are set to offset $\distance{q}[i]$ (the differences introduced by the query variable $q[i]$) in the assertions, a counterexample is found by making all queries fall in the $\true$ branch (i.e., query answers $q[i]$ are all above the threshold $T$). Suppose later, an improved alignment, which properly aligns the branch by $-\distance{q}[i]$, is fed in, which makes the false branch also incur a privacy loss. Therefore a counterexample will then be generated with query answers below the threshold, to make privacy cost exceed the total privacy budget (the last assertion in code).

\subsubsection{Mechanism Generation}\label{sec:util}
In general, mechanism generation runs on both the transformed program $M''$ and the sketch mechanism $M'$ as follows: 
\begin{itemize}
    \item For any candidate solution (of $\hole,\scalehole$) that fails to satisfy any privacy constraint in $M''$ given any previously-generated counterexample, we assign a negative utility score to the solution.
    
    \item Otherwise, we use the utility function $\code{Utility}(M', \hole,\scalehole)$ as its utility score.
\end{itemize}
Based on the utility scores defined above, \tool uses an optimizer to find a privacy mechanism that optimizes the utility function while remaining differentially private.

Returning to our running example. The initial few discovered counterexamples likely include ones that go to different branches to cover all code paths. They can serve as good guides to lead the optimizer towards finding a more general solution, by aligning true and false branch differently, using a conditional alignment in the form of $\ternary{\Omega}{\bullet}{\bullet}$, as other solutions will result in a negative utility score since they violate privacy.

Among the solutions that do satisfy all privacy constraints, the mechanism generation component ranks them based on their utility scores. Here, a solution that assigns a large noise (e.g., $size/\epsilon$) to the queries, although private, will have smaller utility scores than one which assigns $3N/\epsilon$ (since $N < \code{size} / 5$ in precondition). Moreover, a solution that assigns three random variables (two for the threshold, and one for the queries) will be less favorable due to larger sum of variances. This shows the power of our utility metric function in selecting good candidate solutions.

\subsection{Handling While-Private Loop and User-Provided Utility Function}
\label{sec:utility_by_example}

\input{figures/adaptive.tex}

Next, we explore the full-fledged version of \tool, with advanced features of while-private loop and user-provided utility function. We use a recently proposed variant of SVT that we call \code{AdaptiveSVT} (i.e., Adaptive Sparse Vector with Gap in~\cite{freegap}) as an example; its pseudo-code without noise is shown in Figure~\ref{fig:adaptive_base}. Compared with SVT, there are three major changes:
\begin{itemize}
    \item The mechanism uses while-private loop (Line~\ref{line:adaptivesvt:while}) to request the synthesizer to adaptively answer as many quires as possible (the input $N$ specifies the \emph{minimum} number of above-threshold queries that the algorithm should output). 
    
    \item The mechanism partitions query answers into three ranges: $(-\infty,T)$, $[T,T+\sigma)$ and $[T+\sigma,+\infty)$ and requests \tool to automatically allocate the total privacy budget among quires in each range. 

   \item When $q[i]\geq T$, the mechanism releases the gap between $q[i]$ and $T$, instead of a constant.
\end{itemize}
Overall, the mechanism improves over SVT since it can use less privacy budget (i.e., add more noise) for queries that are much larger than the threshold $T$ (i.e., in range $[T+\sigma,+\infty)$), in order to increase the amount of queries that it can process. Moreover, it is shown that the gap information can be released for free~\cite{freegap}.

From program synthesis perspective, it poses two challenges for \tool: (1) to synthesize executable code for while-private loop, and (2) to adopt a user-specified utility function.

\input{figures/adaptive_synthesized}

\paragraph{Synthesizing while-private Loop}
\begin{figure*}
\small
\raggedright
\setstretch{0.85}
\begin{mathpar}
\inferrule*[right=(T-While-Priv)]
{\cdots
}
{\proves \flowrule{\env}{\whilepriv{e}{c}}{\cdots; (\whilecmd{(e\land \vpriv{\priv}\leq \epsilon-\bigcirc)}{(\vpriv{t}=\vpriv{\priv};\cdots;\assert{\vpriv{\priv}-\vpriv{t}\leq \bigcirc}))}}{\env \join \env_f}}
\vspace{-2ex}
\end{mathpar}
\caption{The transformation rule of while-private loop; the parts identical to a standard while-loop are omitted for readability. The complete rule is available in the Appendix.}
\label{fig:trans_while_priv}
\end{figure*}

Recall that in the transformed program $M''$, there is an distinguished variable $\vpriv{\priv}$ that tracks the consumed privacy cost at each program point. The transformation of while-private loop (Figure~\ref{fig:trans_while_priv}) uses $\vpriv{\priv}$ to ensure that the loop terminates if $\vpriv{\priv}$ might exceed $\priv$ after one more iteration: it inserts an unknown bound on the privacy cost of running one iteration ($\bigcirc$) and ensures that the actual cost of \emph{each iteration} never exceeds the bound with the assertion inserted at the end. We note that while-private ($\code{while-priv}$) is a new feature of \tool; it enables \tool to automatically infer and even optimize the loop termination conditions that are previous manually annotated in CheckDP~\cite{checkdp}.

\paragraph*{Discussion on the Soundness of \code{while-priv}}
Although \code{while-priv} is a new feature of \tool, we note that this feature is transformed to a normal \code{while} loop by the transformation rule in Figure~\ref{fig:trans_while_priv}. By construction, the unknown bound on the privacy cost of each loop iteration ($\bigcirc$) is sound. Moreover, as a synthesized mechanism only contains normal \code{while} loops, a synthesized mechanism can further be verified by tools like CheckDP.

\paragraph{User-Specified Utility Function}
Consider the default utility function that minimizes the sum of variances of all random variables (Equation~\ref{eq:default}). A solution that outputs no queries at all always beats other solutions since it injects no noise ($\code{Utility} = \infty$). However, the solution fails the requirement of outputting at least $N$ queries in total, where $N$ is a parameter of the mechanism.
Therefore, a more informative utility function is required for Adaptive SVT. 

Recall that the family of SVTs are designed to report whether a query answer is above a certain threshold or not. Hence, a natural utility measurement is the number of true positives and false positives of the above-threshold queries. Moreover, the design of Adaptive SVT assumes that many queries are well-above the threshold; this allows mechanism to add relatively large noise to the outliers without impacting number of false positives.
Finally, by definition, the synthesized privacy mechanism should output at least $N$ queries in total, where $N$ is a parameter of the mechanism.

Hence, we use a sample input $inp_{ex}$ where many queries are well-above the threshold, create a modified sketch mechanism $M'_\hole$ that removes $\eta_i$ from $M'$ whose alignment is 0, and returns the number of true positives ($\#tp$) and false positives ($\#fp$). Hence, the user-specified utility function is defined as follows:
\begin{equation}
\label{eq:custom_util}
\code{Utility}(M', \hole,\scalehole) = (\#tp - \#fp) - p\times \max(N - (\#tp + \#fp), 0)
\end{equation}
where $p$ is the penalty of outputting less than $N$ outputs, which we set as $1$ to guide the search to favor a solution that answers at least $N$ above-threshold queries.

\paragraph*{Choice of Utility Functions}
The quality of the synthesized mechanism is dependent on the quality of the utility function, as the latter defines ``utility'' in the search. In general, a proper utility function of a privacy mechanism might be both data- and application-specific, such as the data- and application-specific utility function that we derived for Adaptive SVT. Nevertheless, for a variety of mechanisms, as showcased in our evaluation, the default utility function (i.e., the sum of variances of all random variables) already allows \tool to synthesize high quality privacy mechanisms.

%% file: figures/syntax.tex
\begin{figure}
\setstretch{0.9}
\raggedright
\framebox{Syntax of Source Language}
\[
\begin{array}{lccl}
\text{Reals} & r &\in &\mathbb{R} \\
\text{Booleans} & b &\in &\{\true,\false\} \\
\text{Vars} & x  &\in& \nvars \\
\text{Linear Ops} & \oplus &::= &+ \mid - \\
\text{Other Ops} & \otimes &::= &\times \mid / \\
\text{Comparators} & \odot &::= &< \mid > \mid = \mid \leq \mid \geq \\
\text{Bool Exprs} & \bexpr &::=\; &\true \mid \false \mid x \mid \neg \bexpr \mid \nexpr_1 \odot \nexpr_2 \\
\text{Num Exprs} & \nexpr &::=\; &\real \mid x \mid \nexpr_1\oplus \nexpr_2 \mid \nexpr_1\otimes \nexpr_2 \mid \bexpr\mathbin{?}\nexpr_1\mathbin{:}\nexpr_2 \\
\text{Expressions} & e &::=\; &\nexpr \mid  \bexpr \mid e_1::e_2 \mid e_1[e_2]  \\
\text{Commands} & c &::=\; &\skipcmd \mid x := e \mid c_1;c_2 \mid \outcmd e \mid  \\ & & & \ifcmd{e}{(c_1)}{(c_2)} \mid \\
  & & & \whilecmd{e}{(c)} \mid \whilepriv{e}{(c)}\\
\end{array}\vspace{3ex}
\]
\framebox{Syntax of Target Language}
\[
\begin{array}{lccl}
\text{Rand Vars} & \eta &\in& \rvars\\
\text{Num Exprs} & \nexpr &::=\; & \dots \mid \eta \\
\text{Commands} & \synth{c} &::=\; & \skipcmd \mid x := e \mid \synth{c}_1;\synth{c}_2 \mid \outcmd e \mid  \\ & & & \ifcmd{e}{(\synth{c}_1)}{(\synth{c}_2)} \mid \\
  & & & \whilecmd{e}{(\synth{c})} \mid \eta :=\lapm(\nexpr) \\
\end{array}
\]
\caption{\tool: source and target language syntax.}\label{fig:syntax}
\end{figure}

%% file: figures/svt_sketch.tex
\begin{figure}[t]
\raggedright
\noindent\rule{\linewidth}{0.8pt}
\funsigthree{SVT-Sketch}
{\small\code{\priv,T,N,size}:\tyreal,\code{q}:\tylist~\tyreal, \annotation{\lambda:\tylist~\tyreal}}
{(\code{out}: \tylist~\tyreal)}
\algrule
\begin{lstlisting}[frame=none,escapechar=@]
@\annotation{\eta_1 := \lapm{((\scalehole_0 + \scalehole_1 \times N + \scalehole_2 \times T + \scalehole_3 \times size) / \epsilon )}}\label{line:gapsvt_eta_1}@
@\annotation{\replaced{T} := T + \eta_1;}@
i := 0; count := 0;
while (i < size $\land$ count < N)
  @\annotation{\eta_3 := \lapm{((\scalehole_4 + \scalehole_5 \times N + \scalehole_6 \times T + \scalehole_7 \times size) / \epsilon )};}@
  @\annotation{\replaced{T} := \replaced{T} + \eta_3;}@
  @\annotation{\eta_2 := \lapm{((\scalehole_8 + \scalehole_9 \times N + \scalehole_{10} \times T + \scalehole_{11} \times size) / \epsilon )};}@
  @\annotation{\replaced{q} := q[i] + \eta_2;}@
  if (@\annotation{\replaced{q} \geq \replaced{T}}@) then
    out := true::out;
    count := count + 1;
  else
    out := false::out;
  i := i + 1;
\end{lstlisting}
\noindent\rule{\linewidth}{0.8pt}
\caption{Sketch of \code{SVT-Base} with extra noise. \label{fig:svt_sketch}}
\end{figure}

%% file: sec_modeling.tex
\paragraph{Reasoning about Privacy}
To reason about privacy, \tool uses a syntax-directed transformation from the sketch program to non-probabilistic relational code with explicit alignments and proof obligations (i.e., assertions to ensure privacy). For commands, each transformation rule has the following format: 
\[\proves \flowrule{\env}{c}{c'}{\env'}\]
where a typing environment $\env$ tracks for each program variable $x$ its data type with its \emph{distance} written as $\distance{$x$}$. Recall that in the Randomness Alignment technique, the distance of a variable is defined as its value difference across two executions on adjacent query answers (Section~\ref{sec:alignment}). Moreover, $c$ and $c'$ are the sketch code and relational code respectively, and the flow-sensitive type system also updates typing environment to $\env'$ after command $c$.

Most importantly, the transformation inserts assertions to ensure the following (informal) soundness property:
\begin{multline*}
\text{if $M'(inp, \lambda)$ is transformed to
$M''(inp,\distance{$inp$},sample,\hole,\lambda)$, then }\\
\exists \hole, \lambda.~\forall inp,\distance{$inp$},sample.~\textit{all assertions in $M''$ pass}\\
\implies M'(inp,\lambda) \text{ is differentially private}
\end{multline*}

\begin{figure*}
\raggedright
\setstretch{0.85}
\begin{mathpar}
\inferrule*[right=(T-Laplace)]
{\alignment=\code{GenerateTemplate}(\Gamma,\text{All Assertions})\quad c_a = \assert{(\subst{(\eta + \alignment)}{\eta}{\eta_1} = \subst{(\eta + \alignment)}{\eta}{\eta_2} \Rightarrow \eta_1=\eta_2)} } 
{\proves \flowrule{\env}{\eta := \lapm{\scale}}{c_a; \eta := \noise[idx]; idx:= idx+1; \vpriv{\priv} := \vpriv{\priv}+|\alignment|/\scale;\distance{$\eta$} := \alignment ;}{\env[\eta \mapsto \tyreal_*]}}
\end{mathpar}
\caption{A snippet of program transformation rules. $\scale$ represents the scale template instrumented in Phase 1. Distinguished variable $\vpriv{\priv}$ and assertions are added to ensure differential privacy when all assertions are satisfied. The complete transformation rules are available in the~\appendixref.}
\label{fig:trans_sample}
\end{figure*}

The most interesting transformation rule is for the sampling commands in the sketch code, which is shown in Figure~\ref{fig:trans_sample}. 
It performs the following important tasks:
\begin{enumerate}
\item Each sampling command is replaced by a non-probabilistic counterpart ($\eta := sample[idx]; idx:=idx+1;$) that reads a sample from the instrumented function
input $sample$.
\item An alignment template (i.e., $\alignment$) is generated
for each sampling command; each template contains a few
holes, i.e., $\hole$, which is also instrumented as function input. Here, we reuse the \code{GenerateTemplate} function proposed by CheckDP~\cite{checkdp}. Intuitively, the alignments serve as a way to satisfy all inserted assertions in the transformed program. To do so, each alignment template $\alignment_i$ for random variable $\eta_i$ contains distance variables of program variables that (1) appear in assertions, and (2) depend on $\eta_i$. 
Hence, \code{GenerateTemplate} takes the typing environment at the sampling command and all assertions as input, and properly calculates an alignment template, a linear function on a set of relevant distance variables as stated above. 
Since the \code{GenerateTemplate} function is identical to the one used in CheckDP~\cite{checkdp}, we only provide its pseudo-code in the~\appendixref. We refer interested readers to~\cite{checkdp} for a more detailed discussion.
\item The transformed code uses a distinguished variable $\vpriv{\priv}$ to track the overall
privacy cost. Moreover, $\vpriv{\priv}$ is updated to $\vpriv{\priv}+|\alignment|/\scale$, where $\scale$ is the scale template instrumented in Phase 1. As discussed in Section~\ref{sec:alignment}, the update soundly accounts for the privacy cost of aligning the Laplace noise with alignment $\alignment$ and scale $\scale$.
\item 
Assertions are inserted in the transformed code
to ensure the (informal) soundness property stated above.
In particular, it inserts an assertion $c_a$ that checks if the alignment function $\phi(r)=r+\alignment$ is injective (i.e., $\forall r_1, r_2.\phi(r_1)=\phi(r_2)\implies r_1=r_2$). This a fundamental requirement of alignment-based proof~\cite{lightdp}.

\end{enumerate}

For example, the transformed program of the sketch mechanism in Figure~\ref{fig:svt_sketch} is shown in Figure~\ref{fig:svt_trans} with the instrumented code highlighted. Here, each random variable $\eta_i$ is paired with a corresponding alignment template $\alignment_i$ computed by \code{GenerateTemplate}:
\begin{equation}
  \begin{aligned}
    \alignment_1: & \hole_0 \\
  \alignment_2: & (\ternary{\Omega}{\hole_1 + \hole_2 \times
  \distance{$\replaced{T}$} + \hole_3 \times \distance{$q$}[i] }{\hole_4 + \hole_5 \times \distance{$\replaced{T}$} +
  \hole_6 \times \distance{$q$}[i]})   \\\nonumber
  \alignment_3: & (\ternary{\Omega}{\hole_7 + \hole_8 \times
  \distance{$\replaced{T}$} + \hole_9 \times \distance{$q$}[i]}{\hole_{10} + \hole_{11} \times \distance{$\replaced{T}$} +
  \hole_{12} \times \distance{$q$}[i]})  
  \end{aligned}
\end{equation}
where $\Omega$ represents the branch condition at Line~\ref{line:svt_trans:branch}. Note that the privacy cost of each alignment is soundly tracked at Lines 3, 8 and 10. Moreover, the distances of variables (e.g., $\replaced{T}$ and $\replaced{q}$) are properly updated after each assignment. Finally, the transformed code contains assertions to ensure that (1) two related execution of the sketch mechanism will follow the same control flow (e.g., Lines 14 and 18); (2) The distances of output expressions must be zero (not present in Figure~\ref{fig:svt_trans} since the output values are already zero-distance literals; and (3) the overall privacy cost of the program does not exceed the privacy budget (e.g., Line 21) .

\input{figures/svt_transformed.tex}

Since the other transformation rules are mostly identical to those introduced in CheckDP~\cite{checkdp} and the soundness property is a direct implication of Theorem 3 in~\cite{checkdp}, we include the full transformation rules in the Appendix for completeness, and omit the formal statement of the soundness property and its proof in this paper.

\paragraph{Reasoning about Utility}
Note that utility is a property of an instantiation of the mechanism sketch (i.e., fully synthesized program with concrete scales). Hence, reasoning about utility is relatively easy on the mechanism sketch $M'(inp, \lambda)$. The only interesting part is that utility computation should also take into account the alignments $\hole$, as a random variable with $\hole_i=0$ implies that the variable is unnecessary from the privacy perspective; hence, it will be removed in the final synthesized code.

In general, the particular metrics of utility might be application- and data-specific. \tool is designed to be modular: users can plug in their customized utility metrics, and even sample data to optimize the utility of the synthesized privacy mechanism. Hence, in general, \tool is parameterized by a utility function $\code{Utility}(M',\hole,\scalehole)$, where $M'$ is mechanism sketch and $\hole$, $\scalehole$ are the synthesized alignments and scales respectively. By default, \tool uses the sum of variances of all random variables as the utility function (note that \tool currently only supports Laplace noise):\footnote{This is inspired by Lyu et al.~\cite{ninghuisparse} who derived the approximately optimal budget allocation of SVT by minimizing the variance of the branch (Line~\ref{line:svt_branch} in Figure~\ref{fig:svt}).}
\begin{equation}
\label{eq:default}
\code{Utility}(M', \hole,\scalehole) = -(\sum_{\set{\scale_i :\ \alignment_i \neq 0}}2\scale_i^2)
\end{equation}
where $\alignment_i$ and $\scale_i$ denote the synthesized scale and alignment for random variable $\eta_i$. As discussed earlier, we explicitly exclude the ones with $0$ alignments, since they are unnecessary.

Note that to compute utility based on the default utility function, there is no need to execute $M'$. Hence, synthesizing privacy mechanisms with the default utility function is very efficient. Moreover, despite its simplicity, it allows us to synthesize many privacy mechanisms (Section~\ref{sec:evaluation}). For now, we assume the default utility function is in use; how to synthesize with more complicated utility function is deferred to Section~\ref{sec:utility_by_example}.

%% file: figures/svt_transformed.tex
\begin{figure}
\raggedright
\noindent\rule{\linewidth}{0.8pt}
\funsigthree{Transformed SVT}
{\code{T,N,size,q}\instrument{, \code{\distance{q}}, $\noise$, $\hole$, $\scalehole$}}
{(\code{out})}
\algrule
\begin{lstlisting}[frame=none,escapechar=@,firstnumber=1]
@\instrument{$\vpriv{\priv}$ := 0; idx = 0;}@
@\instrument{$\eta_1$ := $\noise$[idx]; idx := idx + 1; \distance{$\eta_1$} := $\alignment_1$;}@
@\instrument{$\vpriv{\priv}$ := $|\alignment_1| / \scale_1$;}@
$\replaced{T}$ := $T + \eta_1$; @\instrument{$\distance{\replaced{T}}$ := $\distance{$\eta_1$}$;}@
count := 0; i := 0;
while (count < N $\land$ i < size)
  @\instrument{$\eta_3$ := $\noise$[idx]; idx := idx + 1; \distance{$\eta_3$} := $\alignment_3$;}@
  @\instrument{$\vpriv{\priv}$ := $\vpriv{\priv} + |\alignment_3| / \scale_3$;}@
  @\instrument{$\eta_2$ := $\noise$[idx]; idx := idx + 1; \distance{$\eta_2$} := $\alignment_2$;}@
  @\instrument{$\vpriv{\priv}$ := $\vpriv{\priv} + |\alignment_2| / \scale_2$;}@
  $\replaced{T}$ := $T + \eta_3$; @\instrument{\distance{$\replaced{T}$} := \distance{$\replaced{T}$} + \distance{$\eta_3$};}@
  $\replaced{q}$ := $q[i] + \eta_2$; @\instrument{\distance{$\replaced{q}$} := \distance{$q$}[i] + \distance{$\eta_2$};}@
  if ($\replaced{q} \geq \replaced{T}$) then@\label{line:svt_trans:branch}@
    @\instrument{\assert{\replaced{q} + \distance{$\replaced{q}$} $\geq$ $\replaced{T} + \distance{$\replaced{T}$}$};}@
    out := $\true$::out;
    count := count + 1;
  else
    @\instrument{\assert{$\lnot$(\replaced{q} + \distance{$\replaced{q}$} $\geq$ $\replaced{T} + \distance{$\replaced{T}$}$)};}@
    out := false::out;
  i := i + 1;
@\instrument{\assert{$\vpriv{\priv} \le \priv$};}@
\end{lstlisting}
\noindent\rule{\linewidth}{0.8pt}
\caption{Transformed mechanism of SVT-Sketch by CheckDP. The instrumented codes are underlined. For better readability, the proof and scale templates are represented by $\alignment_i$ and $\scale_i$, respectively.}
\label{fig:svt_trans}
\end{figure}

%% file: figures/adaptive.tex
\begin{figure}
\raggedright
\noindent\rule{\linewidth}{0.8pt}
\funsigfour{AdaptiveSVT-Base}
{\code{T,N,size,$\sigma$}$: \tyreal$, \code{q}: $\tylist~\synth{\tyreal}$}
{(\code{out}: $\tylist~\tyreal$), \bound{\priv}}
{\alldiffer}
\algrule
\begin{lstlisting}[frame=none,escapechar=@]
i := 0;
while-priv (i < size)@\label{line:adaptivesvt:while}@
  if ($\text{q[i]} - T\geq \sigma$) then
    out := (q[i] - T)::out;
  else
    if ($\text{q[i]}- T \geq 0$) then
      out := (q[i] - T)::out;
    else
      out := 0::out;
  i := i + 1;
\end{lstlisting}
\noindent\rule{\linewidth}{0.8pt}
\caption{AdaptiveSVT-Base, while-private feature is used to enable the synthesis of adaptive mechanisms.\label{fig:adaptive_base}}
\end{figure}

%% file: figures/adaptive_synthesized.tex
\begin{figure}[ht]
\raggedright
\noindent\rule{\linewidth}{0.8pt}
\funsigfour{\small AdaptiveSVT}
{\code{T,N,size,$\sigma$}$:\tyreal$,\code{q}$:\tylist~\tyreal_{*}$}
{(\code{out}:$\tylist~\tyreal$)}
{\alldiffer}
\algrule
\begin{lstlisting}[frame=none,escapechar=@]
i := 0;
$\eta_1$ :=  $\lapm{(4 / \epsilon )}$;
$\vpriv{\priv}$ := $\vpriv{\priv} + 4 / \epsilon$;
$T_{\diamondsuit 1}$ := $T + \eta_1$;
while (i < size $\land$ $\vpriv{\priv} \leq \priv - 2\epsilon / (2N + 3)$)
  $\eta_2$ := $\lapm{((4N + 6) / \epsilon )}$;
  $\vpriv{\priv}$ := $\vpriv{\priv} + (\ternary{\Omega_{Top}}{2}{0}) \times \epsilon / (4N + 6)$;
  $q_{\diamondsuit 1}$ := q[i] + $\eta_2$;
  if ($q_{\diamondsuit 1}$ - $T_{\diamondsuit 1} \geq \sigma$) then
    out := ($q_{\diamondsuit 1}$ - $T_{\diamondsuit 1}$)::out;
  else
    $\eta_3$ := $\lapm{( (2N + 3) / \epsilon )}$;
    $\vpriv{\priv}$ := $\vpriv{\priv} + (\ternary{\Omega_{Middle}}{2}{0}) \times \epsilon / (2N + 3)$;
    $q_{\diamondsuit 2}$ := q[i] + $\eta_3$;
    if ($q_{\diamondsuit 2} - T_{\diamondsuit 1} \geq 0$) then
      out := ($q_{\diamondsuit 2} - T_{\diamondsuit 1}$)::out;
    else
      out := 0::out;
  i := i + 1;
\end{lstlisting}
\noindent\rule{\linewidth}{0.8pt}
\caption{Synthesized AdaptiveSVT based on AdaptiveSVT-Base. $\Omega_{Top}$ and $\Omega_{Middle}$ stands for the branch condition at Line~9 and Line~15, respectively. \label{fig:adaptiev_synthesized}}
\end{figure}

%% file: sec_implementation.tex
We implemented a prototype\footnote{
Publicly available at \url{https://github.com/cmla-psu/dpgen}. }
of \tool in Python. The prototype uses the pyswarms package~\cite{pyswarmsJOSS2018} for PSO optimization. For each component in the CEGIS loop (Figure~\ref{fig:loop}), we run the optimization for 500 iterations. To speedup searching, \tool stops early if the best value stays within tolerance $t = 1$ for 50 iterations. By default, the search space for each hole in the alignments and scales is set to $[-10, 10]$ and $[0, 10]$ respectively. This is chosen based on the typical values of those parameters in correct privacy mechanisms. Moreover, the number of query answers is set to $100$. \tool automatically expands the search space for the holes and the number of query answers until a mechanism is successfully generated. 

We note that the use of the optimizer is to \emph{discover} a solution; the generated mechanism is eventually verified by an off-the-shelf sound verifier CPAChecker~\cite{beyer2011cpachecker} with arbitrary array lengths. Moreover, to speed up the synthesis of adaptive mechanisms, we split the mechanism sketch into multiple sketches each with a unique combinations of different random variable locations, and run all sketches in parallel. To make the generated mechanism easier to read and more friendly for off-the-shelf verifier, we round up the scales of generated mechanism to nearest integer. However, this can be switched off if the user wants a more refined mechanism.

We evaluate \tool on a $\text{Intel}^{\text{\textregistered}}$
$\text{Xeon}^{\text{\textregistered}}$ E5-2620 v4 CPU machine with 64 GB memory. Table~\ref{tab:experiments} lists the synthesized scales, alignments and synthesis time for each mechanism that we introduce next.

\input{tables/experiments.tex}

\subsection{Case Studies}
To illustrate the expressiveness of \tool and its capability of synthesizing privacy mechanisms of different characteristics, we used a standard benchmark as seen in prior works~\cite{shadowdp, checkdp,Ding2018CCS,Bichsel2018CCS,Aws:synthesis}, including SVT under different conditions, other variants of SVT such as NumSVT and GapSVT, the Report Noisy Max mechanism~\cite{diffpbook}, Partial Sum and Smart Sum~\cite{chan10continual}. The psudo-code and transformed program of each case study can be found in the \appendixref. All of the mechanisms we synthesize are proved to satisfy $\priv$-differential privacy. Here we focus on the most interesting mechanisms; the rest can be found in the \appendixref.

\paragraph*{SVT under Different Conditions} As discussed earlier (Section~\ref{sec:challenges}), the SVT-Base program can be made private in multiple ways, and its utility might depend on the characteristics of the data being analyzed as well. 

For example, the standard SVT mechanism makes the use of the fact that the number of above-threshold queries to answer ($N$) is relatively small (hence the name ``sparse vector''). This is specified by the precondition $N < size / 5$ in the function signature. Given this assumption on data, \tool successfully synthesizes the privacy mechanism shown in Figure~\ref{fig:svt}, which is the standard SVT mechanism.

In the SVT-All case, we change the assumption to be most queries answers are above the threshold. Under this assumption, the standard SVT is no longer preferred, as intuitively, the privacy cost paid for the threshold can no longer be offset by its gain from paying no cost for the below-threshold queries.
As expected, \tool synthesizes a privacy mechanism that only injects noise to query answers but not to the threshold, which is the same as the mechanism shown in Figure~\ref{fig:svtalt}.

In the SVT-Inverse case, we flip SVT to answer at most $N$ \emph{below-threshold} queries, rather than
to answer at most $N$ above-threshold queries. Accordingly, the same precondition $N < size / 5$ in the function signature now specifies that the number of \emph{below-threshold} queries to answer is relatively small. 
Not surprisingly, \tool successfully synthesizes the dual of standard SVT, with flipped alignments on the $\true$ and $\false$ branches but the same scales and random variables.

\paragraph*{Finding Approximately Optimal Budget Allocation For SVT} As shown in~\cite{ninghuisparse}, the approximately optimal budget allocation between the threshold and queries for SVT is $1:(1+ (2N)^{\frac{2}{3}})$. Although \tool currently lacks the ability to solve the optimization problem with a symbolic $N$, we analyze a case where input $N$ of SVT is fixed to 1. Also, we disabled integer rounding for synthesizing the approximately optimal allocation for this particular instance of SVT. \tool is able to synthesize a solution with scale $2.587430/\priv$ on $\eta_1$, the noise added to the threshold, and scale $3.259844/\priv$ on $\eta_2$, the noise added to each query answer; while the approximately optimal ones are $2.587401 / \epsilon$ and  $3.259960 /\epsilon$ respectively when $N=1$.

\paragraph*{Variants of SVT Using while-private Loop}
To showcase the power of while-private loop and user-provided utility function, we evaluate on two mechanisms that use these features. The first is Adaptive SVT, which is already introduced in Section~\ref{sec:utility_by_example}. The second, called SVT-WhilePriv, is a modified version of SVT where the user simply uses a while-private loop and asks the synthesizer to adaptively adjust the privacy cost paid to the above and below threshold answers respectively:
\begin{lstlisting}[frame=none,escapechar=@]
i := 0;
while-priv (i < size)
  if ($\text{q[i]} \geq T$) then 
    out := true::out;
  else
    out := false::out;
  i := i + 1;
\end{lstlisting}

In both cases, we use the user-provided utility function of Equation~\ref{eq:custom_util} (Section~\ref{sec:utility_by_example}). The utility function requires the user provide an example input for evaluation. To capture the characteristics of a typical usage of SVT, where the amount of above-threshold query answers is small, we designed an input as follows: we use a sample set of 100 query answers where 75 are well below the threshold ($\leq T-1000$), 10 are well above the threshold ($\geq T+1000$) and 15 close to the threshold ($=T+50$). In both cases, the input $N$ (the minimum number of above-threshold queries to answer) is set to 20 in order to avoid answering queries that are well-below the threshold. 

Moreover, due to the nature of the utility function, which computes utility based on true positives and false positives, we need to run the sketch mechanism (with randomness) many iterations for a good estimation of the utility. In the evaluation, we set the number of iterations to $2500$.

For SVT-WhilePriv, \tool successfully synthesizes a privacy mechanism that is identical to standard SVT: the synthesized mechanism adds noise with scale $3/\epsilon$ (resp. $3N/\epsilon$) to the threshold (resp. each query answer). The synthesized while condition is $\code{\textbf{while}}\ (i < size \land \vpriv{\priv} \le \priv - 2\priv/(3N))$. The synthesized program increments $\vpriv{\priv}$ by $\priv/3$ before the branch, increments it by $2\priv/(3N)$ in the true branch and leaves it unchanged in the false branch (as the alignment in that case is 0). Note that although the synthesized code is syntactically different from standard SVT, they have exactly the same semantics. 

For AdaptiveSVT, \tool synthesizes a version (last row in Table~\ref{tab:experiments}) that is different from the one proposed in~\cite{freegap}. However, we confirmed that the average utility score for the synthesized mechanism across 2500 iterations is $24.4$, meaning that it almost answers all above-threshold queries in an accurate way, with no false positives. In this case, \tool has successfully synthesized a \emph{private} solution that offers better utility (as measured by the user-provided utility function) on the sample data compared with the original mechanism in~\cite{freegap}. In practice, a user could provide more sample data to avoid over-fitting, with the cost of a longer synthesis time.

\paragraph*{Report Noisy Max}

Another well-known privacy mechanism is Report Noisy Max: it finds the \emph{identity} of the item with the maximum score in the database. We present this mechanism in a simplified manner: given a series of query answers as inputs, the mechanism returns the index of the query with maximum answer. 

The synthesis of Report Noisy Max requires an extension of the alignment-based proof technique called Shadow Execution~\cite{shadowdp}, which is also supported by \tool. 
While the synthesis time for Report Noisy Max is slightly longer than other mechanisms without while-private loop, \tool synthesizes a private mechanism that is the same as the standard Report Noisy Max.

\paragraph*{Partial Sum and Smart Sum}

These two mechanisms release aggregate statistics. Partial Sum simply sums up all query answers and directly release the final sum. A more advanced mechanism~\cite{chan10continual} is proposed by Chen et al. to release the prefix sum of a series query answers: $q[0], q[0] + q[1],\cdots, \sum_{i=0}^T q[i]$. The details of this mechanism can be found in the \appendixref.
Notably, these two mechanisms rely on a slightly different adjanceny definitions: at most one of the query answers can differ by at most $1$: 

\centerline{\onediffer}

Despite such difference, \tool is able to synthesizes both Partial Sum and Smart Sum.

\paragraph{Comparison with KOLAHAL~\cite{learning}}
As the implementation of KOLAHAL is not publicly available, we were unable to make a direct comparison with KOLAHAL on all benchmarks. The last column of Table~\ref{tab:experiments} shows data that we collect from~\cite{learning} when the corresponding mechanism is also evaluated on KOLAHAL (N/A is listed if the mechanism is not part of the experiments in~\cite{learning}). We note that one difference between \tool and KOLAHAL is that the latter synthesizes a set of candidate solutions instead of one; sometimes, the ideal solution might not be a top candidate:  
for example, the ideal solution for Smart Sum is ranked as the $4^{\text{th}}$ one. 
Moreover, KOLAHAL requires manually-provided mechanism sketches where noise locations are annotated, while \tool automatically generates sketches as discussed in Section~\ref{sec:sketch}.

\subsection{Performance}

We note that the synthesis time with the default utility function is significantly smaller than the one with a user-provided utility function (used for SVT-WhilePriv and AdaptiveSVT). The reason is that the default utility function does not need to execute the sketch mechanism at all. Moreover, among the ones using the default utility function, Report Noisy Max takes longer to synthesis, since it needs to use the Shadow Execution~\cite{shadowdp} feature of \tool.

\paragraph*{Comparison with KOLAHAL} We note that for the same mechanisms, the synthesis time of \tool is considerably smaller than that of KOLAHAL. While this is not an apple-to-apple comparison, 
we contribute the efficiency to the reduced search space of our sketch generation algorithm and the qualities of the counterexamples generated in the search loop.

%% file: tables/experiments.tex
\newcommand\noisymaxbranch{\ensuremath{\Omega_{NM}}}
\newcommand\svtbranch{\ensuremath{\Omega_{SVT}}}
\newcommand\adaptivetopbranch{\ensuremath{\Omega_{Top}}}
\newcommand\adaptivemiddlebranch{\ensuremath{\Omega_{Middle}}}
\newcommand\tablegap{\arrayrulecolor{lightgray}\hline}

\begin{table*}
\centering
\small{
\caption{Synthesized random variables with corresponding alignment proof. $\Omega_*$ stands for the branch condition in each mechanism, where $\noisymaxbranch = \replaced{q} > bq \lor i = 0$, $\svtbranch = \replaced{q}\ge\replaced{T}$, $\adaptivetopbranch = q_{\diamondsuit 1} - T_{\diamondsuit 1}\ge \sigma$, $\adaptivemiddlebranch = q_{\diamondsuit 2} - T_{\diamondsuit 1} \ge 0$. Unnecessary random variables that are removed in the optimization are omitted.\label{tab:experiments}}
\small
\resizebox{\linewidth}{!}{
\begin{tabular}{ccc|cc|ccc||c}
\Xhline{1.5\arrayrulewidth}
\multirow{3}{*}{\textbf{Mechanism}} & \multicolumn{6}{c}{\textbf{Random Variables}}  & \multirow{3}{*}{\textbf{Time (s)}} & \multirow{3}{*}{\textbf{KOLAHAL~\cite{learning}}} \\
\cline{2-7} & \multicolumn{2}{c}{$\mathbf{\eta_1}$} 
& \multicolumn{2}{c}{$\mathbf{\eta_2}$}  & \multicolumn{2}{c}{$\mathbf{\eta_3}$} & & \\
\cline{2-7} & \textbf{Scale}  &  \textbf{Alignment}
&\textbf{Scale}  &  \textbf{Alignment}  & \textbf{Scale}  &  \textbf{Alignment}  & & \\
\Xhline{1\arrayrulewidth}
\rule{0pt}{1ex}
ReportNoisyMax  &  $2/\epsilon$ & \begin{tabular}{@{}c@{}}
$\ternary{\noisymaxbranch}{1-\distance{$q$}[i]}{0}$\end{tabular}   & N/A & N/A & N/A & N/A  & 120 & 1920  \\ \tablegap
PartialSum & $1 /\epsilon$ & $-\distance{sum}$ & N/A & N/A & N/A & N/A & 10 & 900  \\ \tablegap
SmartSum & $2 /\epsilon$ &  $-\distance{$sum$}-\distance{$q$}[i]$ & $2 /\epsilon$ & $-\distance{$q$}[i]$   & N/A & N/A & 25 & $5460^*$ \\ \tablegap
SVT & $3 /\epsilon$ & 1 &  $3N /\epsilon$ & $\ternary{\Omega_{SVT}}{1 - \distance{q}[i]}{0}$ & N/A & N/A &  29 & 2640 \\ \tablegap
SVT-Inverse & $3/ \epsilon$ & -1 & $3N/\epsilon$ & $\ternary{\Omega_{SVT}}{0}{-2}$ & N/A & N/A  &  28 & N/A  \\\tablegap
SVT-All & N/A & N/A & $\code{size}/\epsilon$ & $\ternary{\Omega_{SVT}}{1}{-1}$ & N/A & N/A  & 38 & N/A \\\tablegap
SVT (N = 1) & $2.587430 /\epsilon$ & 1 & $3.259844 /\epsilon$ & $\ternary{\Omega_{SVT}}{1 - \distance{q}[i]}{0}$ & N/A & N/A &  16 & N/A \\\tablegap
GapSVT & $3 /\epsilon$ & 1 &  $3N /\epsilon$ & $\ternary{\Omega_{SVT}}{1 - \distance{q}[i]}{0}$ & N/A & N/A &  25 & N/A \\\tablegap
NumSVT & $4 /\epsilon$ & 1 & $4N /\epsilon$ & $\ternary{\svtbranch}{2}{0}$ & $4N /\epsilon$ & $-\distance{$q$}[i]$  &  35 & N/A  \\\tablegap
SVT-WhilePriv & $3 /\epsilon$ & 1 & $3N /\epsilon$ & $\ternary{\svtbranch}{2}{0}$ & N/A & N/A  & 617 & N/A  \\\tablegap
AdaptiveSVT & $4/\epsilon$ & 1 & $(4N+6)/\epsilon$ & $\ternary{\adaptivetopbranch}{ 1 - \distance{$q$}[i]}{0}$ & $(2N+3)/\epsilon$ & $\ternary{\adaptivemiddlebranch}{1 - \distance{$q$}[i]}{0}$  &  3026 & N/A \\
\Xhline{1.5\arrayrulewidth}
\end{tabular}}
\footnotesize{$^*$ The ideal solution was ranked $4^{\text{th}}$ among the candidates generated by KOLAHAL.} \hfill
}
\end{table*}

%% file: sec_related.tex
\paragraph*{Synthesizing Differentially Private Algorithms}

Closest to our work is the synthesizer KOLAHAL  recently proposed by Roy et al. ~\cite{learning}. KOLAHAL takes, as inputs, a sketch mechanism with noise expressions in known locations as holes and a finite grammar for noise expressions, and leverages counterexamples generated by StatDP~\cite{Ding2018CCS} and continuous optimization approximation to guide the optimization of noise functions. It supports multiple noise distributions (Laplace, Exponential) and is the first tool capable of synthesizing complex differential privacy mechanisms including NoisyMax, SVT and SmartSum.
Compared with KOLAHAL, \tool (1) automatically generates the locations of randoms variables, (2) is more efficient in synthesizing non-adaptive mechanisms due to reduced search space of the templates, and (3) is able to synthesize sophisticated mechanisms such as AdaptiveSVT.
An earlier synthesizer~\cite{synthesisdp} relies on user supplied examples and uses a sensitivity-directed program synthesis technique based on DFuzz~\cite{DFuzz}. However, it can only synthesize simple mechanisms where the privacy analysis follows directly from the composition theorem.

\paragraph*{Proving and Disproving Differential Privacy} Differential privacy has been a fruitful target for formal verification due to its compositional property. 
Fuzz \cite{Fuzz} and DFuzz \cite{DFuzz} use linear dependent type systems to analyse program sensitivity and prove (pure) differential privacy properties. Amorim et al. \cite{ExtFuzz} extend such systems to work under approximate differential privacy. %
Barthe et al. \cite{Barthe12,EasyCrypt,BartheICALP2013,Barthe16,BartheCCS16} developed several customized relational logics based on probabilistic couplings for reasoning about differential privacy. Zhang and Kifer \cite{lightdp} introduced the Randomness Alignment technique as a simpler but more restricted alternative of probabilistic coupling. Wang et al. \cite{shadowdp} extend the type system in \cite{lightdp} to allow more complicated Randomness Alignment functions to be used for sophisticated mechanisms. Albarghouthi and
Hsu \cite{Aws:synthesis} synthesize probabilistic couplings and randomness alignment into coupling strategies, creating the first fully automated tool capable of generating coupling proofs for complex mechanisms.

A complementary line of work \cite{Ding2018CCS,Bichsel2018CCS, dpsniper} is concerned with developing automated techniques to search for counterexamples that witness violations of differential privacy. StatDP \cite{Ding2018CCS} uses statistical hypothesis testing to demonstrate high probability of privacy violations. 
DP-Finder \cite{Bichsel2018CCS} uses symbolic differentiation and gradient descent to search for counterexamples. 
More recently, DP-Sniper \cite{dpsniper} trains a classifier -- a parametric family of posterior probability distributions to predict if an observed output is likely generated from one of two possible inputs, and use this classifier to select a set of outputs that can best distinguish these two inputs.  
All these methods rely on \emph{sampling} -- running an algorithm hundreds of thousands of times to estimate the output distribution of mechanisms and generate counterexample candidates/training data. 

Recent work \cite{checkdp,barthe2020, FarinaCG2021} targets both proving and disproving differential privacy. CheckDP~\cite{checkdp} also relies on the Randomness Alignment technique. It reduces the search space of proofs to templates with holes. Moreover, it embeds a novel bi-directional CEGIS loop to improve proof and counterexample simultaneously.
Barthe et al.~\cite{barthe2020} identify a non-trivial class of programs where checking (pure and approximate) differential privacy is decidable. However, these programs only allow a bounded number of samples from the Laplace distribution, and their inputs and outputs are from a finite domain.
Farina~\cite{FarinaCG2021} builds a relational symbolic execution framework, which when combined with probabilistic couplings, is able to prove differential privacy for SVT or generate failing traces for its two incorrect variants. 

%% file: sec_conclusion.tex
In this paper, we present \tool, an automated differential privacy mechanism synthesizer that is able to synthesize sophisticated DP mechanisms such as adaptive mechanisms. \tool employs a novel approach to automatically generate sketch mechanisms with potential random variables, and uses an enhanced CEGIS loop to fill the holes in the sketch according to customizable utility functions. Compared with recent synthesis work, \tool is reasonably faster in synthesizing non-adaptive mechanisms, and is the only tool that is powerful enough to synthesize sophisticated adaptive ones. Evaluations show \tool synthesizes a variety of non-adaptive mechanisms within minutes and adaptive ones within an hour. 

Future work includes exploring more utility metrics for optimizing mechanism, as well as extending \tool to support solving the optimization problem symbolically, which provides more general forms of budget allocations to different random variables in the mechanism. Another possibility is to extend the underlying proof technique (i.e., randomness alignment) to support more complex mechanisms such as PrivTree, where the intermediate results depend on the data, but the aggregate result does not. Moreover, we focus on Laplace distribution due to its adoption in a variety of mechanisms as shown in our benchmark. In general, new random distributions can be added to alignment-based proofs in a modular way via extra typing rules, as showcased in~\cite{lightdp}. Extending DPGen for other random distributions is another potential future direction.

%% file: sec_appendix.tex
\section{Full Case Studies}

In this section we list the examples we studied in this paper. For each mechanism we show the original mechanism (the user's input), and the transformed mechanism for the synthesis loop.

\input{figures/noisymax.tex}

\input{figures/numsvt.tex}

\input{figures/gapsvt.tex}

\input{figures/svt_inverse.tex}
\clearpage

\input{figures/partialsum.tex}

  \section{Pseudo-code for \code{GenerateTemplate}}
  Here for completeness, we include the pseudo-code of the helper function \code{GenerateTemplate} proposed by~\cite{checkdp}. Note that \code{Depends} is a variable dependence checking oracle which returns \true\ if the expression $e$ depends on the variable $\eta$. This oracle can be implemented as standard program dependency analysis~\cite{aho1986compilers,ferrante1987} or information flow analysis~\cite{Bergeretti:1985:IDA:2363.2366}.
  
  \begin{algorithm}[ht]
  \setstretch{0.9}
  \SetKwProg{Fn}{function}{\string:}{}
  \SetKwFunction{Depends}{Depends}
  \SetKw{Break}{break}
  \SetKwFunction{GenerateTemplate}{GenerateTemplate}
  \SetKwInOut{Input}{input}
  \DontPrintSemicolon
  \Input{ $\env_s$: typing environment at sampling command \\ 
  $A$: set of the generated assertions in the program} 
  \Fn{\GenerateTemplate{$\env_s$, $A$}}{
  $\exprset \gets \emptyset$, $\varset \gets \emptyset$\;
  \ForEach{$\assert{e} \in A$}{
      \If{$\Depends(e, \eta)$}{\label{line:template_depends_eta}
          \If{$\assert{e}$ is generated by \ruleref{T-If}}{
              $e' \gets $ the branch condition of \code{\textbf{if}}\;
                  $\exprset \gets \exprset \cup \{e'\}$\;
              
          }
          \ForEach{$v \in Vars\cup \{e_1[e_2] | e_1[e_2] \in e\}$ }{
              \If{$\Gamma_s\not\proves v:\basety_0 \land \Depends(e, v)$}{
                  $\varset \gets \varset \cup \{v\}$ \;
              }
          }
      }
  }
  \ForEach{$e \in \exprset\cup \varset$ }{
  remove $e$ from \exprset and \varset if not in scope\;
  }
  \Return \exprset,\varset;
  }
  \caption{Template generation for $\eta := \lapm{r}$}
  \end{algorithm}

\input{figures/smartsum.tex}

\section{Complete Transformation Rules}
In this section we list the transformation rules in Figure~\ref{fig:trans_rules} for completeness. Note that most rules are identical to the ones in CheckDP~\cite{checkdp}, with the differences highlighted in gray.

\input{figures/complete_rules.tex}

%% file: figures/noisymax.tex
\begin{figure}[H]
\raggedright
\noindent\rule{\linewidth}{2\arrayrulewidth}
\funsigfour{NoisyMax}
{\code{size}$:\tyreal$, \code{q}$:\tylist~\synth{\tyreal}$}
{\code{max}$:\tyreal$}
{$\forall$ \code{i}.~ $-1\leq$~\code{$\distance{\text{q}}$[i]}~$\leq1$}
\algrule
\begin{lstlisting}[frame=none, escapechar=@,basicstyle=\appendixalgsize\ttfamily]
i := 0; bq := 0; max := 0;
while (i < size)
  if (q[i] > bq $\lor$ i = 0)@\label{line:noisymax_branch}@
   max := i;@\label{line:noisymax_true_branch}@@\label{line:noisymax_out}@
   bq := q[i];@\label{line:noisymax_true_branch_2}@
  i := i + 1;
\end{lstlisting}
\noindent\rule{\linewidth}{0.8pt}
\funsigtwo{\small Transformed NoisyMax}
{\code{size,q}\instrument{, \distance{\code{q}}, $\noise$, $\hole$, $\scalehole$}}
\algrule
\begin{lstlisting}[frame=none, escapechar=@,firstnumber=8,basicstyle=\appendixalgsize\ttfamily]
@\instrument{$\vpriv{\epsilon}$ := 0; idx := 0;}@
i := 0; bq := 0; max := 0;
@\instrument{$\eta_2$ := $\noise$[idx]; idx := idx + 1; \distance{$\eta_2$} := $\alignment_2$;}@
@\instrument{$\vpriv{\epsilon}$ := $\vpriv{\priv}$ + $|\alignment_2| / \scale_2$;}@
$\replaced{bq}$ := bq + $\eta_2$;
@\instrument{\distance{$\replaced{bq}$} := \distance{$\eta_2$};}@
$\instrument{\first{\distance{\code{bq}}}\text{ := 0;}\quad\second{\distance{\code{bq}}}\text{ := 0;}\quad\first{\distance{\code{max}}} \text{ := 0;}\quad\second{\distance{\code{max}}}\text{ := 0;}}$ /*\quad\second{\distance{\code{max}}}\text{ := 0;}*/
while (i < size)
  @\instrument{$\eta_1$ := $\noise$[idx]; idx := idx + 1;  \distance{$\eta_1$} := $\alignment_1$;}@
  @\instrument{$\vpriv{\epsilon}$ := ($\ternary{\mathcal{L}_1}{\vpriv{\epsilon}}{0}$) + $|\alignment_1| / \scale_1$;}@
  $\replaced{q}$ := q[i] + $\eta_1$;
  @\instrument{\distance{$\replaced{q}$} := \distance{$\eta_1$}}@;
  @\instrument{$\eta_3$ := $\noise$[idx]; idx := idx + 1;  \distance{$\eta_3$} := $\alignment_3$;}@
  @\instrument{$\vpriv{\epsilon}$ := ($\ternary{\mathcal{L}_2}{\vpriv{\epsilon}}{0}$) + $|\alignment_3| / \scale_3$;}@
  $\replaced{bq}$ := bq + $\eta_3$;
  @\instrument{\distance{$\replaced{bq}$} := \distance{$\replaced{bq}$} + \distance{$\eta_3$};}@
  @\instrument{\code{\textbf{if}} ($\mathcal{L}_1$)\quad$\first{\distance{bq}}$ := $\second{\distance{bq}}$; $\first{\distance{max}}$ := $\second{\distance{max}}$;}@
  @\instrument{\code{\textbf{if}} ($\mathcal{L}_2$)\quad$\first{\distance{bq}}$ := $\second{\distance{bq}}$; $\first{\distance{max}}$ := $\second{\distance{max}}$;}@
  if ($\replaced{q}$ > $\replaced{bq}$ $\lor$ i = 0) @\label{line:noisymax:branch}@
    @\instrument{\assert{$\text{q[i]}+\distance{\text{q}}\text{[i]}+\eta+\first{\eta} > \text{bq} + \first{\text{bq}} \lor \text{i} = 0$};}@
    max := i;
    @\instrument{$\first{\code{max}}$ := 0;}@
    $\instrument{\second{\distance{\text{bq}}}\text{ := bq + }\second{\distance{\text{bq}}}\text{ - (q[i] + }\eta\text{);}}$@\label{line:noisymax_bq_shadow}@
    bq := q[i] + $\eta$;
    @\instrument{$\first{\distance{\text{bq}}}$ := $\first{\distance{\text{q}}}$[i] + $\first{\distance{$\eta$}}$;}@
  $\instrument{\code{\textbf{else}}}$
    @\instrument{\assert{$\lnot(\text{q[i]}+\distance{\text{q}}\text{[i]}+\eta+\first{\eta} > \text{bq} + \first{\text{bq}} \lor \text{i} = 0)$};}@  
  // shadow execution@\label{line:noisymax_shadow_exe_start}@
  $\instrument{\code{\textbf{if}}~(\text{q[i]} + \second{\distance{\text{q}}}\text{[i]} + \eta> \text{bq} + \second{\distance{\text{bq}}}\mathbin{\lor}\text{i = 0})}$@\label{line:noisymax_shadow_condition}@
    $\instrument{\second{\distance{\text{bq}}}\text{ := q[i] + }\second{\distance{\text{q}}}\text{[i]} + \eta - \text{bq}\text{;}}$
    @\instrument{$\second{\distance{\text{max}}}$ := i - max;}@@\label{line:noisymax_shadow_exe_end}@
  i := i + 1;
@\instrument{\assert{$\vpriv{\priv} \le \priv$};}@
\end{lstlisting}
\noindent\rule{\linewidth}{2\arrayrulewidth}
\caption{Report Noisy Max and its transformed code.$\mathcal{L}_i$ stands for shadow execution selectors. }
\end{figure}

%% file: figures/numsvt.tex
\begin{figure}[H]
\raggedright
\small
\noindent\rule{\linewidth}{0.8pt}
\funsigfour{NumSVT}
{\code{T,N,size}: \tyreal, \code{q}: \tylist~$\synth{\tyreal}$}
{(\code{out}: \tylist~\tyreal), \bound{\priv}}
{\alldiffer $\land$ \code{N < \code{size / 5}}}
\algrule
\begin{lstlisting}[frame=none,escapechar=@,basicstyle=\appendixalgsize\ttfamily]
count := 0; i := 0;
while (count < N $\land$ i < size)
  if ($\text{q[i]}\geq T$) then
    out := (q[i])::out;
    count := count + 1;
  else
    out := false::out;
  i := i + 1;
\end{lstlisting}
\noindent\rule{\linewidth}{0.8pt}
\funsigthree{Transformed NumSVT}
{\code{T,N,size,q}\instrument{, \code{\distance{q}}, $\noise$, $\hole$, $\scalehole$}}
{(\code{out})}
\algrule
\begin{lstlisting}[frame=none,escapechar=@,firstnumber=1]
@\instrument{$\vpriv{\priv}$ := 0; idx = 0;}@
@\instrument{$\eta_1$ := $\noise$[idx]; idx := idx + 1; \distance{$\eta_1$} := $\alignment_1$;}@
@\instrument{$\vpriv{\priv}$ := $|\alignment_1| / \scale_1$;}@
$\replaced{T}$ := $T + \eta_1$; @\instrument{$\distance{\replaced{T}}$ := $\distance{$\eta_1$}$;}@
count := 0; i := 0;
while (count < N $\land$ i < size)
  @\instrument{$\eta_4$ := $\noise$[idx]; idx := idx + 1; \distance{$\eta_4$} := $\alignment_4$;}@
  @\instrument{$\vpriv{\priv}$ := $\vpriv{\priv} + |\alignment_4| / \scale_4$;}@
  $\replaced{T}$ := $T + \eta_4$; @\instrument{\distance{$\replaced{T}$} := \distance{$\replaced{T}$} + \distance{$\eta_4$};}@
  @\instrument{$\eta_2$ := $\noise$[idx]; idx := idx + 1; \distance{$\eta_2$} := $\alignment_2$;}@
  @\instrument{$\vpriv{\priv}$ := $\vpriv{\priv} + |\alignment_2| / \scale_2$;}@
  $\replaced{q}$ := $q[i] + \eta_2$; @\instrument{\distance{$\replaced{q}$} := \distance{$q$}[i] + \distance{$\eta_2$};}@
  if ($\replaced{q} \geq \replaced{T}$) then
    @\instrument{$\eta_3$ := $\noise$[idx]; idx := idx + 1; \distance{$\eta_3$} := $\alignment_3$;}@
    @\instrument{$\vpriv{\priv}$ := $\vpriv{\priv} + |\alignment_3| / \scale_3$;}@
    $\replaced{q}$ := $q[i] + \eta_3$; @\instrument{\distance{$\replaced{q}$} := \distance{$q$}[i] + \distance{$\eta_3$};}@
    @\instrument{\assert{\replaced{q} + \distance{$\replaced{q}$} $\geq$ $\replaced{T} + \distance{$\replaced{T}$}$};}@
    out := ($\replaced{q}$)::out;
    count := count + 1;
  else
    @\instrument{\assert{$\lnot$(\replaced{q} + \distance{$\replaced{q}$} $\geq$ $\replaced{T} + \distance{$\replaced{T}$}$)};}@
    out := false::out;
  i := i + 1;
@\instrument{\assert{$\vpriv{\priv} \le \priv$};}@
\end{lstlisting}
\noindent\rule{\linewidth}{0.8pt}
\caption{Numerical Sparse Vector Technique and its transformed code. \label{fig:numsvt}}
\end{figure}

%% file: figures/gapsvt.tex
\begin{figure}
\raggedright
\noindent\rule{\linewidth}{0.8pt}
\funsigfour{GapSVT-Base}
{\code{T,N,size}: \tyreal, \code{q}: \tylist~$\synth{\tyreal}$}
{(\code{out}: \tylist~\tyreal), \bound{\priv}}
{\alldiffer $\land$ \code{N < \code{size / 5}}}
\algrule
\begin{lstlisting}[frame=none,escapechar=@]
i := 0; count := 0;
while (i < size $\land$ count < N)
  if ($\text{q[i]} \geq T$) then 
    out := (q[i] - T)::out;
  else
    out := false::out;
    count := count + 1;
  i := i + 1;
\end{lstlisting}
\noindent\rule{\linewidth}{0.8pt}
\funsigthree{Transformed GapSVT}
{\code{T,N,size,q}\instrument{, \code{\distance{q}}, $\noise$, $\hole$, $\scalehole$}}
{(\code{out})}
\algrule
\begin{lstlisting}[frame=none,escapechar=@,firstnumber=1]
@\instrument{$\vpriv{\priv}$ := 0; idx = 0;}@
@\instrument{$\eta_1$ := $\noise$[idx]; idx := idx + 1; \distance{$\eta_1$} := $\alignment_1$;}@
@\instrument{$\vpriv{\priv}$ := $|\alignment_1| / \scale_1$;}@
$\replaced{T}$ := $T + \eta_1$; @\instrument{$\distance{\replaced{T}}$ := $\distance{$\eta_1$}$;}@
count := 0; i := 0;
while (count < N $\land$ i < size)
  @\instrument{$\eta_2$ := $\noise$[idx]; idx := idx + 1; \distance{$\eta_2$} := $\alignment_2$;}@
  @\instrument{$\vpriv{\priv}$ := $\vpriv{\priv} + |\alignment_2| / \scale_2$;}@
  @\instrument{$\eta_3$ := $\noise$[idx]; idx := idx + 1; \distance{$\eta_3$} := $\alignment_3$;}@
  @\instrument{$\vpriv{\priv}$ := $\vpriv{\priv} + |\alignment_3| / \scale_3$;}@
  $\replaced{T}$ := $T + \eta_2$; @\instrument{\distance{$\replaced{T}$} := \distance{$\replaced{T}$} + \distance{$\eta_2$};}@
  $\replaced{q}$ := $q[i] + \eta_3$; @\instrument{\distance{$\replaced{q}$} := \distance{$q$}[i] + \distance{$\eta_3$};}@
  if ($\replaced{q} \geq \replaced{T}$) then
    @\instrument{\assert{\replaced{q} + \distance{$\replaced{q}$} $\geq$ $\replaced{T} + \distance{$\replaced{T}$}$};}@
    @\instrument{\assert{\distance{\replaced{q}} - \distance{\replaced{T}} = 0};}@
    out := ($\replaced{q} - \replaced{T}$)::out;
  else
    @\instrument{\assert{$\lnot$(\replaced{q} + \distance{$\replaced{q}$} $\geq$ $\replaced{T} + \distance{$\replaced{T}$}$)};}@
    out := false::out;
    count := count + 1;
  i := i + 1;
@\instrument{\assert{$\vpriv{\priv} \le \priv$};}@
\end{lstlisting}
\noindent\rule{\linewidth}{0.8pt}
\caption{GapSVT and its transformed code.}
\end{figure}

%% file: figures/svt_inverse.tex
\begin{figure}[ht]
\raggedright
\noindent\rule{\linewidth}{0.8pt}
\funsigfour{SVTBase-Inverse}
{\code{T,N,size}: \tyreal, \code{q}: \tylist~$\synth{\tyreal}$}
{(\code{out}: \tylist~\tyreal), \bound{\priv}}
{\alldiffer $\land$ \code{N < \code{size / 5}}}
\algrule
\begin{lstlisting}[frame=none,escapechar=@]
i := 0; count := 0;
while (i < size $\land$ count < N)
  if ($\text{q[i]} \geq T$) then 
    out := true::out;
  else
    out := false::out;
    count := count + 1;
  i := i + 1;
\end{lstlisting}
\noindent\rule{\linewidth}{0.8pt}
\funsigthree{Transformed SVT}
{\code{T,N,size,q}\instrument{, \code{\distance{q}}, $\noise$, $\hole$, $\scalehole$}}
{(\code{out})}
\algrule
\begin{lstlisting}[frame=none,escapechar=@,firstnumber=1]
@\instrument{$\vpriv{\priv}$ := 0; idx = 0;}@
@\instrument{$\eta_1$ := $\noise$[idx]; idx := idx + 1; \distance{$\eta_1$} := $\alignment_1$;}@
@\instrument{$\vpriv{\priv}$ := $|\alignment_1| / \scale_1$;}@
$\replaced{T}$ := $T + \eta_1$; @\instrument{$\distance{\replaced{T}}$ := $\distance{$\eta_1$}$;}@
count := 0; i := 0;
while (count < N $\land$ i < size)
  @\instrument{$\eta_2$ := $\noise$[idx]; idx := idx + 1; \distance{$\eta_2$} := $\alignment_2$;}@
  @\instrument{$\vpriv{\priv}$ := $\vpriv{\priv} + |\alignment_2| / \scale_2$;}@
  @\instrument{$\eta_3$ := $\noise$[idx]; idx := idx + 1; \distance{$\eta_3$} := $\alignment_3$;}@
  @\instrument{$\vpriv{\priv}$ := $\vpriv{\priv} + |\alignment_3| / \scale_3$;}@
  $\replaced{T}$ := $T + \eta_2$; @\instrument{\distance{$\replaced{T}$} := \distance{$\replaced{T}$} + \distance{$\eta_2$};}@
  $\replaced{q}$ := $q[i] + \eta_3$; @\instrument{\distance{$\replaced{q}$} := \distance{$q$}[i] + \distance{$\eta_3$};}@
  if ($\replaced{q} \geq \replaced{T}$) then
    @\instrument{\assert{\replaced{q} + \distance{$\replaced{q}$} $\geq$ $\replaced{T} + \distance{$\replaced{T}$}$};}@
    out := q[i]::out;
  else
    @\instrument{\assert{$\lnot$(\replaced{q} + \distance{$\replaced{q}$} $\geq$ $\replaced{T} + \distance{$\replaced{T}$}$)};}@
    out := false::out;
    count := count + 1;
  i := i + 1;
@\instrument{\assert{$\vpriv{\priv} \le \priv$};}@
\end{lstlisting}
\noindent\rule{\linewidth}{0.8pt}
\caption{SVT-Inverse and its transformed code.}
\end{figure}

%% file: figures/partialsum.tex
\begin{figure}[H]
\raggedright
\setstretch{0.9}
\noindent\rule{\linewidth}{2\arrayrulewidth}
\funsigfour{PartialSum}
{\code{size}$:\tyreal$, \code{q}$:\tylist~\synth{\tyreal}$}
{(\code{out}:\tyreal), \bound{\priv}}
{\onediffer}
\algrule
\begin{lstlisting}[frame=none,escapechar=@]
sum := 0; i := 0;
while (i < size)
  sum := sum + q[i];
  i := i + 1;
out := sum;
\end{lstlisting}
\noindent\rule{\linewidth}{0.8pt}
\funsigthree{Transformed PartialSum}
{\code{size,q}\instrument{,\code{\distance{q}}, $\noise$, $\hole$, $\scalehole$}}
{(\code{out})}
\algrule
\begin{lstlisting}[frame=none,escapechar=@,firstnumber=7]
@\instrument{$\vpriv{\priv}$ := 0;}@
$\replaced{sum}$ := 0; i := 0;
@\instrument{\distance{$\replaced{sum}$} := 0;}@
while (i < size)
  $\replaced{sum}$ := $\replaced{sum}$ + q[i];
  @\instrument{\distance{$\replaced{sum}$} := \distance{$\replaced{sum}$} + \distance{q}[i];}@
  i := i + 1;
@\instrument{$\eta_1$ := $\noise$[idx]; idx := idx + 1; \distance{$\eta_1$} := $\alignment_1$;}@
@\instrument{$\vpriv{\priv}$ := $\vpriv{\priv} + |\alignment_1| / \scale_1$;}@
$\replaced{sum}$ := $\replaced{sum}$ + $\eta_1$; @\instrument{\distance{$\replaced{sum}$} := \distance{$\replaced{sum}$} + \distance{$\eta_1$};}@
@\instrument{\assert{\distance{$\replaced{sum}$} == 0};}@
out := $\replaced{sum}$;
@\instrument{\assert{$\vpriv{\priv} \le \priv$};}@
\end{lstlisting}
\noindent\rule{\linewidth}{2\arrayrulewidth}
\caption{PartialSum and its transformed code.}
\label{fig:partialsum}
\end{figure}

%% file: figures/smartsum.tex
\begin{figure}[H]
\raggedright
\small
\noindent\rule{\linewidth}{2\arrayrulewidth}
\funsigfour{SmartSum}
{\code{M,T,size}$:\tyreal$, \code{q}$:\tylist~\synth{\tyreal}$}
{(\code{out}:\tylist~\tyreal), \bound{$\priv$}}
{\onediffer}
\algrule
\begin{lstlisting}[frame=none, escapechar=@]
i := 0; next := 0; sum := 0;
while (i < size $\land$ i $\leq$ T)
  if ((i + 1) mod M = 0) then 
    next := sum + q[i];
    sum := 0;
    out := next::out; 
  else @\label{line:smartsum_else}@
    next:= next + q[i];
    sum := sum + q[i];
    out := next::out;
  i := i + 1;
\end{lstlisting}
\noindent\rule{\linewidth}{0.8pt}
\funsigthree{Transformed SmartSum}
{\code{M,T,size,q}\instrument{, \code{\distance{q}}, $\noise$, $\hole$, $\scalehole$}}
{(\code{out})}
\algrule
\begin{lstlisting}[frame=none, escapechar=@,firstnumber=14]
@\instrument{$\vpriv{\priv}$ := 0; idx := 0;}@
i := 0; $\replaced{next}$ := 0; $\replaced{sum}$ := 0;
@\instrument{\distance{\replaced{sum}} := 0; \distance{\replaced{next}} := 0;}@
while (i < size $\land$ i $\leq$ T)
  if ((i + 1) mod M = 0) then
    @\instrument{$\eta_1$ := $\noise$[idx]; idx := idx + 1;}@
    @\instrument{$\vpriv{\priv}$ := $\vpriv{\priv}$ + |$\alignment_1$| / $\scale_1$; \distance{$\eta_1$} := $\alignment_1$;}@
    $\replaced{next}$ := $\replaced{sum}$ + q[i] + $\eta_1$; 
    @\instrument{\distance{\replaced{next}} := \distance{$\replaced{sum}$} + \distance{q}[i] + \distance{$\eta_1$};}@
    $\replaced{sum}$ := 0; @\instrument{\distance{$\replaced{sum}$} := 0;}@
    @\instrument{\assert{\distance{\replaced{next}} = 0};}@
    out := next::out;
  else
    @\instrument{$\eta_2$ := $\noise$[idx]; idx := idx + 1;}@
    @\instrument{$\vpriv{\priv}$ := $\vpriv{\priv}$ + |$\alignment_2$| / $\scale_2$; \distance{$\eta_2$} := $\alignment_2$;}@
    $\replaced{next}$ := $\replaced{next}$ + q[i] + $\eta_2$;
    @\instrument{\distance{\replaced{next}} := \distance{\replaced{next}} + \distance{q}[i] + \distance{$\eta_2$};}@
    @\instrument{$\eta_3$ := $\noise$[idx]; idx := idx + 1;}@
    @\instrument{$\vpriv{\priv}$ := $\vpriv{\priv}$ + |$\alignment_3$| / $\scale_3$; \distance{$\eta_3$} := $\alignment_3$;}@
    $\replaced{sum}$ := $\replaced{sum}$ + q[i] + $\eta_3$; 
    @\instrument{\distance{\replaced{sum}} := \distance{\replaced{sum}} + \distance{q}[i] + $\distance{$\eta_3$}$;}@
    @\instrument{\assert{\distance{\replaced{next}} = 0};}@
    out := next::out;
  i := i + 1;
@\instrument{\assert{$\vpriv{\priv} \le \priv$};}@
\end{lstlisting}
\noindent\rule{\linewidth}{2\arrayrulewidth}
\caption{SmartSum and its transformed code.}
\label{alg:smartsum}
\end{figure}

%% file: figures/complete_rules.tex
\begin{figure*}[ht]
\small
\raggedright
\framebox{\textbf{Transformation rules for expressions with form $\env \proves e:\basety_{\nexpr}$}}
\begin{mathpar}
  \inferrule*[right=(T-Num)]{ }
  { \env \proves \exprrule{\real}{\tyreal_0}{\real}{\noconstraints}}
  \and
  \inferrule*[right=(T-Boolean)]{ }
  {\env \proves \exprrule{b}{\bool}{b}{\noconstraints}}
  \and
  \inferrule*[right=(T-VarZero)]{ }
  { \env, x:\basety_0 \proves \exprrule{x}{\basety_0}{x}{\noconstraints}}
  \and
  \inferrule*[right=(T-VarStar)]
  { }
  {\env, x:\basety_{*}\proves \exprrule{x}{\basety_{\distance{$x$}}}{x + \distance{$x$}}{\noconstraints}}
  \and
  \inferrule*[right=(T-Neg)]
  {\env\proves \exprrule{e}{\bool}{e'}{\constraints}}
  { \env \proves \exprrule{\neg e}{\bool}{e'}{\constraints}}
  \and
  \inferrule*[right=(T-OPlus)]
  {\env\proves \exprrule{e_1}{\basety_{\nexpr_1}}{e_1'}{\constraints_1} \quad \env\proves \exprrule{e_2}{\basety_{\nexpr_2}}{e_2'}{\constraints_2}{}}
  {\env \proves \exprrule{e_1 \oplus e_2}{\basety_{\nexpr_1 \oplus \nexpr_2}}{e_1' \oplus e_2'}{\constraints_1 \land \constraints_2}}
  \and
  \inferrule*[right=(T-OTimes)]
  {\env\proves \exprrule{e_1}{\tyreal_{\nexpr_1}}{}{\constraints_1} \quad \env\proves \exprrule{e_2}{\tyreal_{\nexpr_2}}{}{\constraints_2}}
  {\env \proves \exprrule{e_1 \otimes e_2}{\tyreal_0}{}{\constraints_1 \land \constraints_2 \land (\nexpr_1 = \nexpr_2 = 0)}}
  \quad
  \inferrule*[right=(T-ODot)]
  {\env\proves \exprrule{e_1}{\tyreal_{\nexpr_1}}{}{\constraints_1} \quad \env\proves \exprrule{e_1}{\tyreal_{\nexpr_2}}{}{\constraints_2}}
  {\env \proves \exprrule{e_1 \odot e_2}{\bool}{}{\constraints_1 \land \constraints_2 \land \inferrule{}{(e_1 \odot e_2) \Leftrightarrow \\\\ (e_1 + \nexpr_1) \odot (e_2 + \nexpr_2)}}}
  \and
  \inferrule*[right=(T-Cons)]
  {\env\proves \exprrule{e_1}{\basety_{\nexpr_1}}{}{\constraints_1} \quad \env\proves \exprrule{e_2}{\tylist~\basety_{\nexpr_2}}{}{\constraints_2}}
  {\env \proves \exprrule{e_1::e_2}{\tylist~\basety_{\nexpr}}{}{\constraints_1 \land \constraints_2 \land (\nexpr_1 = \nexpr_2 = 0)}}
  \and
  \inferrule*[right=(T-Index)]
  {\env\proves \exprrule{e_1}{\tylist~\tau}{}{\constraints_1}\quad \env\proves \exprrule{e_2}{\tyreal_{\nexpr}}{}{\constraints_2}}
  {\env \proves \exprrule{e_1[e_2]}{\tau}{}{\constraints_1 \land \constraints_2 \land (\nexpr = 0)}}
  \and
  \inferrule*[right=(T-Select)]
  {\env\proves \exprrule{e_1}{\bool}{}{\constraints_1} \quad \env\proves \exprrule{e_2}{\basety_{\nexpr_1}}{}{\constraints_2} \quad \env\proves \exprrule{e_3}{\basety_{\nexpr_2}}{}{\constraints_3}}
  {\env \proves \exprrule{e_1\mathbin{?}e_2\mathbin{:}e_3}{\basety_{\nexpr_1}}{}{\constraints_1 \land \constraints_2 \land \constraints_3\land (\nexpr_1=\nexpr_2)}}
  \end{mathpar}
\framebox{\textbf{Transformation rules for commands with form $\proves \flowrule{\env}{c}{c'}{\env'}$}}
\begin{mathpar}

\inferrule*[right=(T-Asgn)]
{ \env \proves \exprrule{e}{\basety_{\nexpr}}{}{\constraints} \quad \configtwo{\dexpr}{c} = \begin{cases}
\configtwo{0}{\skipcmd} , \ \text{ if } \nexpr == 0, \cr
\configtwo{*}{\distance{$x$} := \nexpr} , \text{ otherwise} 
\end{cases}}
{\proves \flowrule{\env}{x := e;}{\assert{\constraints}; x := e; c}{\env[x\mapsto \basety_{\dexpr}]}}
\and
\inferrule*[right=(T-Seq)]
{\proves \flowrule{\env}{c_1}{c_1'}{\env_1} \quad \proves \flowrule{\env_1}{c_2}{c_2'}{\env_2}}
{\proves\flowrule{\env}{c_1;c_2}{c_1';c_2'}{\env_2}}
\and
\inferrule*[right=(T-Return)]
{ \env \proves \exprrule{e}{\basety_{\nexpr}}{}{\constraints}}
{\proves \flowrule{\env}{\outcmd{e}}{\assert{\constraints\land \nexpr = 0};\outcmd{e}}{\env}}
\and
\inferrule*[right=(T-Skip)]{ }
{\proves \flowrule{\env}{\skipcmd}{\skipcmd}{\env}}
\and
\inferrule*[right=(T-While)]
{\proves\flowrule{\env \join \env_f}{c}{c'}{\env_f} \quad \env, \env \join \env_f \Rrightarrow c_s \quad \env_f, \env \join \env_f \Rrightarrow c''
}
{\proves \flowrule{\env}{\whilecmd{e}{c}}{c_s; (\whilecmd{e}{(\assert{\aligndexec{e, \env}};c';c''))}}{\env \join \env_f}}
\and
\annotation{\inferrule*[right=(T-While-Priv)]
{\proves\flowrule{\env \join \env_f}{c}{c'}{\env_f} \quad \env, \env \join \env_f \Rrightarrow c_s \quad \env_f, \env \join \env_f \Rrightarrow c''
}
{\proves \flowrule{\env}{\whilepriv{e}{c}}{c_s; (\whilecmd{(e\land \vpriv{\priv}\leq \epsilon-\bigcirc)}{(\assert{\aligndexec{e, \env}};\vpriv{t}=\vpriv{\priv};c';c'';\assert{\vpriv{\priv}-\vpriv{t}\leq \bigcirc}))}}{\env \join \env_f}}}
\and
\inferrule*[right=(T-If)]
{
\proves \flowrule{\env}{c_i}{c_i'}{\env_i} \quad \env_i, \env_1 \join \env_2 \Rrightarrow c''_i \quad i \in \{1, 2\} 
}
{\proves \flowrule{\env}{\ifcmd{e}{c_1}{c_2}}{\ifcmd{e}{(\assert{\aligndexec{e, \env}};c_1';c_1'')}{(\assert{\lnot\aligndexec{e, \env}};c_2';c_2'')}}{\env_1 \join \env_2}}
\and
\inferrule*[right=(T-Laplace)]
{\alignment=\code{GenerateTemplate}(\Gamma,\text{All Assertions})\quad c_a = \assert{(\subst{(\eta + \alignment)}{\eta}{\eta_1} = \subst{(\eta + \alignment)}{\eta}{\eta_2} \Rightarrow \eta_1=\eta_2)} } 
{\proves \flowrule{\env}{\eta := \lapm{\annotation{\scale}}}{c_a; \eta := \noise[idx]; idx:= idx+1; \vpriv{\priv} := \vpriv{\priv}+|\alignment|\annotation{/\scale};\distance{$\eta$} := \alignment ;}{\env[\eta \mapsto \tyreal_*]}}
\end{mathpar}
\framebox{\textbf{Transformation rules for merging environments}}
\begin{mathpar}
\inferrule*
{\env_1 \sqsubseteq \env_2 \quad c= \{\distance{$x$} := 0 \mid \env_1(x)=\tyreal_0 \land \env_2(x) = \tyreal_*\}}
{\env_1, \env_2 \Rrightarrow c}
\end{mathpar}
\caption{Program transformation rules. $\scale$ represents the scale template instrumented in Phase 1. Distinguished variable $\vpriv{\priv}$ and assertions are added to ensure differential privacy.\label{fig:trans_rules}}
\end{figure*}

%% file: main.bbl

\begin{thebibliography}{49}


\ifx \showCODEN    \undefined \def \showCODEN     #1{\unskip}     \fi
\ifx \showDOI      \undefined \def \showDOI       #1{#1}\fi
\ifx \showISBNx    \undefined \def \showISBNx     #1{\unskip}     \fi
\ifx \showISBNxiii \undefined \def \showISBNxiii  #1{\unskip}     \fi
\ifx \showISSN     \undefined \def \showISSN      #1{\unskip}     \fi
\ifx \showLCCN     \undefined \def \showLCCN      #1{\unskip}     \fi
\ifx \shownote     \undefined \def \shownote      #1{#1}          \fi
\ifx \showarticletitle \undefined \def \showarticletitle #1{#1}   \fi
\ifx \showURL      \undefined \def \showURL       {\relax}        \fi
\providecommand\bibfield[2]{#2}
\providecommand\bibinfo[2]{#2}
\providecommand\natexlab[1]{#1}
\providecommand\showeprint[2][]{arXiv:#2}

\bibitem[\protect\citeauthoryear{Abowd}{Abowd}{2018}]%
        {abowd18kdd}
\bibfield{author}{\bibinfo{person}{John~M. Abowd}.}
  \bibinfo{year}{2018}\natexlab{}.
\newblock \showarticletitle{The U.S. Census Bureau Adopts Differential
  Privacy}. In \bibinfo{booktitle}{\emph{Proceedings of the 24th ACM SIGKDD
  International Conference on Knowledge Discovery and Data Mining}} (London,
  United Kingdom) \emph{(\bibinfo{series}{KDD '18})}. \bibinfo{publisher}{ACM},
  \bibinfo{address}{New York, NY, USA}, \bibinfo{pages}{2867--2867}.
\newblock
\showISBNx{978-1-4503-5552-0}


\bibitem[\protect\citeauthoryear{Aho, Sethi, and Ullman}{Aho
  et~al\mbox{.}}{1986}]%
        {aho1986compilers}
\bibfield{author}{\bibinfo{person}{Alfred~V Aho}, \bibinfo{person}{Ravi Sethi},
  {and} \bibinfo{person}{Jeffrey~D Ullman}.} \bibinfo{year}{1986}\natexlab{}.
\newblock \showarticletitle{Compilers, principles, techniques}.
\newblock \bibinfo{journal}{\emph{Addison wesley}} \bibinfo{volume}{7},
  \bibinfo{number}{8} (\bibinfo{year}{1986}), \bibinfo{pages}{9}.
\newblock


\bibitem[\protect\citeauthoryear{Albarghouthi and Hsu}{Albarghouthi and
  Hsu}{2017}]%
        {Aws:synthesis}
\bibfield{author}{\bibinfo{person}{Aws Albarghouthi} {and}
  \bibinfo{person}{Justin Hsu}.} \bibinfo{year}{2017}\natexlab{}.
\newblock \showarticletitle{Synthesizing Coupling Proofs of Differential
  Privacy}.
\newblock \bibinfo{journal}{\emph{Proceedings of ACM Programming Languages}}
  \bibinfo{volume}{2}, \bibinfo{number}{POPL}, Article \bibinfo{articleno}{58}
  (\bibinfo{date}{dec} \bibinfo{year}{2017}), \bibinfo{numpages}{30}~pages.
\newblock
\showISSN{2475-1421}


\bibitem[\protect\citeauthoryear{Barthe, Chadha, Jagannath, Sistla, and
  Viswanathan}{Barthe et~al\mbox{.}}{2020}]%
        {barthe2020}
\bibfield{author}{\bibinfo{person}{Gilles Barthe}, \bibinfo{person}{Rohit
  Chadha}, \bibinfo{person}{Vishal Jagannath}, \bibinfo{person}{A.~Prasad
  Sistla}, {and} \bibinfo{person}{Mahesh Viswanathan}.}
  \bibinfo{year}{2020}\natexlab{}.
\newblock \showarticletitle{Deciding Differential Privacy for Programs with
  Finite Inputs and Outputs}. In \bibinfo{booktitle}{\emph{Proceedings of the
  35th Annual ACM/IEEE Symposium on Logic in Computer Science}}
  (Saarbr\"{u}cken, Germany) \emph{(\bibinfo{series}{LICS ’20})}.
  \bibinfo{publisher}{Association for Computing Machinery},
  \bibinfo{address}{New York, NY, USA}, \bibinfo{pages}{141–154}.
\newblock
\showISBNx{9781450371049}
\urldef\tempurl%
\url{https://doi.org/10.1145/3373718.3394796}
\showDOI{\tempurl}


\bibitem[\protect\citeauthoryear{Barthe, Danezis, Gregoire, Kunz, and
  Zanella-Beguelin}{Barthe et~al\mbox{.}}{2013}]%
        {EasyCrypt}
\bibfield{author}{\bibinfo{person}{Gilles Barthe}, \bibinfo{person}{George
  Danezis}, \bibinfo{person}{Benjamin Gregoire}, \bibinfo{person}{Cesar Kunz},
  {and} \bibinfo{person}{Santiago Zanella-Beguelin}.}
  \bibinfo{year}{2013}\natexlab{}.
\newblock \showarticletitle{Verified Computational Differential Privacy with
  Applications to Smart Metering}. In \bibinfo{booktitle}{\emph{Proceedings of
  the 2013 IEEE 26th Computer Security Foundations Symposium}}
  \emph{(\bibinfo{series}{CSF '13})}. \bibinfo{publisher}{IEEE Computer
  Society}, \bibinfo{address}{Washington, DC, USA}, \bibinfo{pages}{287--301}.
\newblock
\showISBNx{978-0-7695-5031-2}


\bibitem[\protect\citeauthoryear{Barthe, Fong, Gaboardi, Gr{\'e}goire, Hsu, and
  Strub}{Barthe et~al\mbox{.}}{2016a}]%
        {BartheCCS16}
\bibfield{author}{\bibinfo{person}{Gilles Barthe}, \bibinfo{person}{No{\'e}mie
  Fong}, \bibinfo{person}{Marco Gaboardi}, \bibinfo{person}{Benjamin
  Gr{\'e}goire}, \bibinfo{person}{Justin Hsu}, {and}
  \bibinfo{person}{Pierre-Yves Strub}.} \bibinfo{year}{2016}\natexlab{a}.
\newblock \showarticletitle{Advanced Probabilistic Couplings for Differential
  Privacy}. In \bibinfo{booktitle}{\emph{Proceedings of the 2016 ACM SIGSAC
  Conference on Computer and Communications Security}} (Vienna, Austria)
  \emph{(\bibinfo{series}{CCS '16})}. \bibinfo{publisher}{ACM},
  \bibinfo{address}{New York, NY, USA}, \bibinfo{pages}{55--67}.
\newblock
\showISBNx{978-1-4503-4139-4}


\bibitem[\protect\citeauthoryear{Barthe, Gaboardi, Gr{\'e}goire, Hsu, and
  Strub}{Barthe et~al\mbox{.}}{2016b}]%
        {Barthe16}
\bibfield{author}{\bibinfo{person}{Gilles Barthe}, \bibinfo{person}{Marco
  Gaboardi}, \bibinfo{person}{Benjamin Gr{\'e}goire}, \bibinfo{person}{Justin
  Hsu}, {and} \bibinfo{person}{Pierre-Yves Strub}.}
  \bibinfo{year}{2016}\natexlab{b}.
\newblock \showarticletitle{Proving Differential Privacy via Probabilistic
  Couplings}. In \bibinfo{booktitle}{\emph{Proceedings of the 31st Annual
  ACM/IEEE Symposium on Logic in Computer Science}} (New York, NY, USA)
  \emph{(\bibinfo{series}{LICS '16})}. \bibinfo{publisher}{ACM},
  \bibinfo{address}{New York, NY, USA}, \bibinfo{pages}{749--758}.
\newblock
\showISBNx{978-1-4503-4391-6}


\bibitem[\protect\citeauthoryear{Barthe, K\"{o}pf, Olmedo, and
  Zanella~B{\'e}guelin}{Barthe et~al\mbox{.}}{2012}]%
        {Barthe12}
\bibfield{author}{\bibinfo{person}{Gilles Barthe}, \bibinfo{person}{Boris
  K\"{o}pf}, \bibinfo{person}{Federico Olmedo}, {and} \bibinfo{person}{Santiago
  Zanella~B{\'e}guelin}.} \bibinfo{year}{2012}\natexlab{}.
\newblock \showarticletitle{Probabilistic Relational Reasoning for Differential
  Privacy}. In \bibinfo{booktitle}{\emph{Proceedings of the 39th Annual ACM
  SIGPLAN-SIGACT Symposium on Principles of Programming Languages}}
  (Philadelphia, PA, USA) \emph{(\bibinfo{series}{POPL '12})}.
  \bibinfo{publisher}{ACM}, \bibinfo{address}{New York, NY, USA},
  \bibinfo{pages}{97--110}.
\newblock
\showISBNx{978-1-4503-1083-3}


\bibitem[\protect\citeauthoryear{Barthe and Olmedo}{Barthe and Olmedo}{2013}]%
        {BartheICALP2013}
\bibfield{author}{\bibinfo{person}{Gilles Barthe} {and}
  \bibinfo{person}{Federico Olmedo}.} \bibinfo{year}{2013}\natexlab{}.
\newblock \showarticletitle{Beyond Differential Privacy: Composition Theorems
  and Relational Logic for f-divergences Between Probabilistic Programs}. In
  \bibinfo{booktitle}{\emph{Proceedings of the 40th International Conference on
  Automata, Languages, and Programming - Volume Part II}} (Riga, Latvia)
  \emph{(\bibinfo{series}{ICALP'13})}. \bibinfo{publisher}{Springer-Verlag},
  \bibinfo{address}{Berlin, Heidelberg}, \bibinfo{pages}{49--60}.
\newblock
\showISBNx{978-3-642-39211-5}


\bibitem[\protect\citeauthoryear{Bergeretti and Carr{\'e}}{Bergeretti and
  Carr{\'e}}{1985}]%
        {Bergeretti:1985:IDA:2363.2366}
\bibfield{author}{\bibinfo{person}{Jean-Francois Bergeretti} {and}
  \bibinfo{person}{Bernard~A. Carr{\'e}}.} \bibinfo{year}{1985}\natexlab{}.
\newblock \showarticletitle{Information-flow and Data-flow Analysis of
  While-programs}.
\newblock \bibinfo{journal}{\emph{ACM Trans. Program. Lang. Syst.}}
  \bibinfo{volume}{7}, \bibinfo{number}{1} (\bibinfo{date}{Jan.}
  \bibinfo{year}{1985}), \bibinfo{pages}{37--61}.
\newblock
\showISSN{0164-0925}
\urldef\tempurl%
\url{https://doi.org/10.1145/2363.2366}
\showDOI{\tempurl}


\bibitem[\protect\citeauthoryear{Beyer and Keremoglu}{Beyer and
  Keremoglu}{2011}]%
        {beyer2011cpachecker}
\bibfield{author}{\bibinfo{person}{Dirk Beyer} {and} \bibinfo{person}{M.~Erkan
  Keremoglu}.} \bibinfo{year}{2011}\natexlab{}.
\newblock \showarticletitle{CPACHECKER: A Tool for Configurable Software
  Verification}. In \bibinfo{booktitle}{\emph{Proceedings of the 23rd
  International Conference on Computer Aided Verification}} (Snowbird, UT)
  \emph{(\bibinfo{series}{CAV'11})}. \bibinfo{publisher}{Springer-Verlag},
  \bibinfo{address}{Berlin, Heidelberg}, \bibinfo{pages}{184--190}.
\newblock
\showISBNx{978-3-642-22109-5}


\bibitem[\protect\citeauthoryear{Bichsel, Gehr, Drachsler-Cohen, Tsankov, and
  Vechev}{Bichsel et~al\mbox{.}}{2018}]%
        {Bichsel2018CCS}
\bibfield{author}{\bibinfo{person}{Benjamin Bichsel}, \bibinfo{person}{Timon
  Gehr}, \bibinfo{person}{Dana Drachsler-Cohen}, \bibinfo{person}{Petar
  Tsankov}, {and} \bibinfo{person}{Martin Vechev}.}
  \bibinfo{year}{2018}\natexlab{}.
\newblock \showarticletitle{DP-Finder: Finding Differential Privacy Violations
  by Sampling and Optimization}. In \bibinfo{booktitle}{\emph{Proceedings of
  the 2018 ACM SIGSAC Conference on Computer and Communications Security}}
  (Toronto, Canada) \emph{(\bibinfo{series}{CCS '18})}.
  \bibinfo{publisher}{ACM}, \bibinfo{address}{New York, NY, USA},
  \bibinfo{pages}{508--524}.
\newblock
\showISBNx{978-1-4503-5693-0}


\bibitem[\protect\citeauthoryear{Bichsel, Steffen, Bogunovic, and
  Vechev}{Bichsel et~al\mbox{.}}{2021}]%
        {dpsniper}
\bibfield{author}{\bibinfo{person}{B. Bichsel}, \bibinfo{person}{S. Steffen},
  \bibinfo{person}{I. Bogunovic}, {and} \bibinfo{person}{M. Vechev}.}
  \bibinfo{year}{2021}\natexlab{}.
\newblock \showarticletitle{DP-Sniper: Black-Box Discovery of Differential
  Privacy Violations using Classifiers}. In \bibinfo{booktitle}{\emph{2021 2021
  IEEE Symposium on Security and Privacy (SP)}}. \bibinfo{publisher}{IEEE
  Computer Society}, \bibinfo{address}{Los Alamitos, CA, USA},
  \bibinfo{pages}{391--409}.
\newblock
\showISSN{2375-1207}
\urldef\tempurl%
\url{https://doi.org/10.1109/SP40001.2021.00081}
\showDOI{\tempurl}


\bibitem[\protect\citeauthoryear{Bittau, Erlingsson, Maniatis, Mironov,
  Raghunathan, Lie, Rudominer, Kode, Tinnes, and Seefeld}{Bittau
  et~al\mbox{.}}{2017}]%
        {prochlo}
\bibfield{author}{\bibinfo{person}{Andrea Bittau}, \bibinfo{person}{\'{U}lfar
  Erlingsson}, \bibinfo{person}{Petros Maniatis}, \bibinfo{person}{Ilya
  Mironov}, \bibinfo{person}{Ananth Raghunathan}, \bibinfo{person}{David Lie},
  \bibinfo{person}{Mitch Rudominer}, \bibinfo{person}{Ushasree Kode},
  \bibinfo{person}{Julien Tinnes}, {and} \bibinfo{person}{Bernhard Seefeld}.}
  \bibinfo{year}{2017}\natexlab{}.
\newblock \showarticletitle{Prochlo: Strong Privacy for Analytics in the
  Crowd}. In \bibinfo{booktitle}{\emph{Proceedings of the 26th Symposium on
  Operating Systems Principles}} (Shanghai, China) \emph{(\bibinfo{series}{SOSP
  '17})}. \bibinfo{publisher}{ACM}, \bibinfo{address}{New York, NY, USA},
  \bibinfo{pages}{441--459}.
\newblock
\showISBNx{978-1-4503-5085-3}
\urldef\tempurl%
\url{https://doi.org/10.1145/3132747.3132769}
\showDOI{\tempurl}


\bibitem[\protect\citeauthoryear{Bureau}{Bureau}{2019}]%
        {onthemap}
\bibfield{author}{\bibinfo{person}{U.~S.~Census Bureau}.}
  \bibinfo{year}{2019}\natexlab{}.
\newblock \bibinfo{title}{On The Map: Longitudinal Employer-Household
  Dynamics}.
\newblock
  \bibinfo{howpublished}{\url{https://lehd.ces.census.gov/applications/help/onthemap.html\#!confidentiality_protection}}.
\newblock


\bibitem[\protect\citeauthoryear{Chan, Shi, and Song}{Chan
  et~al\mbox{.}}{2011}]%
        {chan10continual}
\bibfield{author}{\bibinfo{person}{T.-H.~Hubert Chan}, \bibinfo{person}{Elaine
  Shi}, {and} \bibinfo{person}{Dawn Song}.} \bibinfo{year}{2011}\natexlab{}.
\newblock \showarticletitle{Private and Continual Release of Statistics}.
\newblock \bibinfo{journal}{\emph{ACM Trans. Inf. Syst. Secur.}}
  \bibinfo{volume}{14}, \bibinfo{number}{3}, Article \bibinfo{articleno}{26}
  (\bibinfo{date}{Nov.} \bibinfo{year}{2011}), \bibinfo{numpages}{24}~pages.
\newblock
\showISSN{1094-9224}


\bibitem[\protect\citeauthoryear{Chen and Machanavajjhala}{Chen and
  Machanavajjhala}{2015}]%
        {ashwinsparse}
\bibfield{author}{\bibinfo{person}{Yan Chen} {and} \bibinfo{person}{Ashwin
  Machanavajjhala}.} \bibinfo{year}{2015}\natexlab{}.
\newblock \bibinfo{title}{On the Privacy Properties of Variants on the Sparse
  Vector Technique}.
\newblock \bibinfo{howpublished}{http://arxiv.org/abs/1508.07306}.
\newblock


\bibitem[\protect\citeauthoryear{de~Amorim, Gaboardi, Hsu, and
  Katsumata}{de~Amorim et~al\mbox{.}}{2019}]%
        {ExtFuzz}
\bibfield{author}{\bibinfo{person}{Arthur~Azevedo de Amorim},
  \bibinfo{person}{Marco Gaboardi}, \bibinfo{person}{Justin Hsu}, {and}
  \bibinfo{person}{Shin-ya Katsumata}.} \bibinfo{year}{2019}\natexlab{}.
\newblock \bibinfo{booktitle}{\emph{Probabilistic Relational Reasoning via
  Metrics}}.
\newblock \bibinfo{publisher}{IEEE Press}.
\newblock


\bibitem[\protect\citeauthoryear{Ding, Kulkarni, and Yekhanin}{Ding
  et~al\mbox{.}}{2017}]%
        {DingKY17}
\bibfield{author}{\bibinfo{person}{Bolin Ding}, \bibinfo{person}{Janardhan
  Kulkarni}, {and} \bibinfo{person}{Sergey Yekhanin}.}
  \bibinfo{year}{2017}\natexlab{}.
\newblock \showarticletitle{Collecting Telemetry Data Privately}. In
  \bibinfo{booktitle}{\emph{Proceedings of the 31st International Conference on
  Neural Information Processing Systems}} (Long Beach, California, USA)
  \emph{(\bibinfo{series}{NIPS'17})}. \bibinfo{publisher}{Curran Associates
  Inc.}, \bibinfo{address}{USA}, \bibinfo{pages}{3574--3583}.
\newblock
\showISBNx{978-1-5108-6096-4}
\urldef\tempurl%
\url{http://dl.acm.org/citation.cfm?id=3294996.3295115}
\showURL{%
\tempurl}


\bibitem[\protect\citeauthoryear{Ding, Wang, Wang, Zhang, and Kifer}{Ding
  et~al\mbox{.}}{2018}]%
        {Ding2018CCS}
\bibfield{author}{\bibinfo{person}{Zeyu Ding}, \bibinfo{person}{Yuxin Wang},
  \bibinfo{person}{Guanhong Wang}, \bibinfo{person}{Danfeng Zhang}, {and}
  \bibinfo{person}{Daniel Kifer}.} \bibinfo{year}{2018}\natexlab{}.
\newblock \showarticletitle{Detecting Violations of Differential Privacy}. In
  \bibinfo{booktitle}{\emph{Proceedings of the 2018 ACM SIGSAC Conference on
  Computer and Communications Security}} (Toronto, Canada)
  \emph{(\bibinfo{series}{CCS \'18})}. \bibinfo{publisher}{ACM},
  \bibinfo{address}{New York, NY, USA}, \bibinfo{pages}{475--489}.
\newblock
\showISBNx{978-1-4503-5693-0}


\bibitem[\protect\citeauthoryear{Ding, Wang, Zhang, and Kifer}{Ding
  et~al\mbox{.}}{2019}]%
        {freegap}
\bibfield{author}{\bibinfo{person}{Zeyu Ding}, \bibinfo{person}{Yuxin Wang},
  \bibinfo{person}{Danfeng Zhang}, {and} \bibinfo{person}{Daniel Kifer}.}
  \bibinfo{year}{2019}\natexlab{}.
\newblock \showarticletitle{Free Gap Information from the Differentially
  Private Sparse Vector and Noisy Max Mechanisms}.
\newblock \bibinfo{journal}{\emph{PVLDB}} \bibinfo{volume}{13},
  \bibinfo{number}{3} (\bibinfo{year}{2019}), \bibinfo{pages}{293--306}.
\newblock
\urldef\tempurl%
\url{https://doi.org/10.14778/3368289.3368295}
\showDOI{\tempurl}


\bibitem[\protect\citeauthoryear{Dwork}{Dwork}{2006}]%
        {Dwork06diffpriv}
\bibfield{author}{\bibinfo{person}{Cynthia Dwork}.}
  \bibinfo{year}{2006}\natexlab{}.
\newblock \showarticletitle{Differential Privacy}. In
  \bibinfo{booktitle}{\emph{Proceedings of the 33rd International Conference on
  Automata, Languages and Programming - Volume Part II}} (Venice, Italy)
  \emph{(\bibinfo{series}{ICALP'06})}. \bibinfo{publisher}{Springer-Verlag},
  \bibinfo{address}{Berlin, Heidelberg}, \bibinfo{pages}{1--12}.
\newblock
\showISBNx{3-540-35907-9, 978-3-540-35907-4}
\urldef\tempurl%
\url{https://doi.org/10.1007/11787006_1}
\showDOI{\tempurl}


\bibitem[\protect\citeauthoryear{Dwork, McSherry, Nissim, and Smith}{Dwork
  et~al\mbox{.}}{2006}]%
        {dwork06Calibrating}
\bibfield{author}{\bibinfo{person}{Cynthia Dwork}, \bibinfo{person}{Frank
  McSherry}, \bibinfo{person}{Kobbi Nissim}, {and} \bibinfo{person}{Adam
  Smith}.} \bibinfo{year}{2006}\natexlab{}.
\newblock \showarticletitle{Calibrating Noise to Sensitivity in Private Data
  Analysis}. In \bibinfo{booktitle}{\emph{Theory of Cryptography}},
  \bibfield{editor}{\bibinfo{person}{Shai Halevi} {and} \bibinfo{person}{Tal
  Rabin}} (Eds.). \bibinfo{publisher}{Springer Berlin Heidelberg},
  \bibinfo{address}{Berlin, Heidelberg}, \bibinfo{pages}{265--284}.
\newblock
\showISBNx{978-3-540-32732-5}


\bibitem[\protect\citeauthoryear{Dwork, Roth, et~al\mbox{.}}{Dwork
  et~al\mbox{.}}{2014}]%
        {diffpbook}
\bibfield{author}{\bibinfo{person}{Cynthia Dwork}, \bibinfo{person}{Aaron
  Roth}, {et~al\mbox{.}}} \bibinfo{year}{2014}\natexlab{}.
\newblock \showarticletitle{The algorithmic foundations of differential
  privacy}.
\newblock \bibinfo{journal}{\emph{Theoretical Computer Science}}
  \bibinfo{volume}{9}, \bibinfo{number}{3--4} (\bibinfo{year}{2014}),
  \bibinfo{pages}{211--407}.
\newblock


\bibitem[\protect\citeauthoryear{Erlingsson, Pihur, and Korolova}{Erlingsson
  et~al\mbox{.}}{2014}]%
        {rappor}
\bibfield{author}{\bibinfo{person}{\'{U}lfar Erlingsson},
  \bibinfo{person}{Vasyl Pihur}, {and} \bibinfo{person}{Aleksandra Korolova}.}
  \bibinfo{year}{2014}\natexlab{}.
\newblock \showarticletitle{RAPPOR: Randomized Aggregatable Privacy-Preserving
  Ordinal Response}. In \bibinfo{booktitle}{\emph{Proceedings of the 2014 ACM
  SIGSAC Conference on Computer and Communications Security}} (Scottsdale,
  Arizona, USA) \emph{(\bibinfo{series}{CCS '14})}. \bibinfo{publisher}{ACM},
  \bibinfo{address}{New York, NY, USA}, \bibinfo{pages}{1054--1067}.
\newblock
\showISBNx{978-1-4503-2957-6}


\bibitem[\protect\citeauthoryear{Farina, Chong, and Gaboardi}{Farina
  et~al\mbox{.}}{2021}]%
        {FarinaCG2021}
\bibfield{author}{\bibinfo{person}{Gian~Pietro Farina},
  \bibinfo{person}{Stephen Chong}, {and} \bibinfo{person}{Marco Gaboardi}.}
  \bibinfo{year}{2021}\natexlab{}.
\newblock \showarticletitle{Coupled Relational Symbolic Execution for
  Differential Privacy}.
\newblock \bibinfo{journal}{\emph{Programming Languages and Systems}}
  \bibinfo{volume}{12648} (\bibinfo{year}{2021}), \bibinfo{pages}{207}.
\newblock


\bibitem[\protect\citeauthoryear{Ferrante, Ottenstein, and Warren}{Ferrante
  et~al\mbox{.}}{1987}]%
        {ferrante1987}
\bibfield{author}{\bibinfo{person}{Jeanne Ferrante}, \bibinfo{person}{Karl~J
  Ottenstein}, {and} \bibinfo{person}{Joe~D Warren}.}
  \bibinfo{year}{1987}\natexlab{}.
\newblock \showarticletitle{The program dependence graph and its use in
  optimization}.
\newblock \bibinfo{journal}{\emph{ACM Transactions on Programming Languages and
  Systems (TOPLAS)}} \bibinfo{volume}{9}, \bibinfo{number}{3}
  (\bibinfo{year}{1987}), \bibinfo{pages}{319--349}.
\newblock


\bibitem[\protect\citeauthoryear{Gaboardi, Haeberlen, Hsu, Narayan, and
  Pierce}{Gaboardi et~al\mbox{.}}{2013}]%
        {DFuzz}
\bibfield{author}{\bibinfo{person}{Marco Gaboardi}, \bibinfo{person}{Andreas
  Haeberlen}, \bibinfo{person}{Justin Hsu}, \bibinfo{person}{Arjun Narayan},
  {and} \bibinfo{person}{Benjamin~C. Pierce}.} \bibinfo{year}{2013}\natexlab{}.
\newblock \showarticletitle{Linear Dependent Types for Differential Privacy}.
  In \bibinfo{booktitle}{\emph{Proceedings of the 40th Annual ACM
  SIGPLAN-SIGACT Symposium on Principles of Programming Languages}} (Rome,
  Italy) \emph{(\bibinfo{series}{POPL '13})}. \bibinfo{publisher}{ACM},
  \bibinfo{address}{New York, NY, USA}, \bibinfo{pages}{357--370}.
\newblock
\showISBNx{978-1-4503-1832-7}
\urldef\tempurl%
\url{https://doi.org/10.1145/2429069.2429113}
\showDOI{\tempurl}


\bibitem[\protect\citeauthoryear{Gilbert and McMillan}{Gilbert and
  McMillan}{2018}]%
        {diffproptest}
\bibfield{author}{\bibinfo{person}{Anna Gilbert} {and} \bibinfo{person}{Audra
  McMillan}.} \bibinfo{year}{2018}\natexlab{}.
\newblock \bibinfo{title}{Property Testing for Differential Privacy}.
\newblock
\newblock
\showeprint[arxiv]{1806.06427}~[cs.CR]


\bibitem[\protect\citeauthoryear{Goguen and Meseguer}{Goguen and
  Meseguer}{1982}]%
        {noninterference}
\bibfield{author}{\bibinfo{person}{J.~A. Goguen} {and} \bibinfo{person}{J.
  Meseguer}.} \bibinfo{year}{1982}\natexlab{}.
\newblock \showarticletitle{Security Policies and Security Models}. In
  \bibinfo{booktitle}{\emph{1982 IEEE Symposium on Security and Privacy}}.
  \bibinfo{publisher}{IEEE}, \bibinfo{address}{Los Alamitos, CA, USA},
  \bibinfo{pages}{11--11}.
\newblock
\urldef\tempurl%
\url{https://doi.org/10.1109/SP.1982.10014}
\showDOI{\tempurl}


\bibitem[\protect\citeauthoryear{Gulwani, Jha, Tiwari, and Venkatesan}{Gulwani
  et~al\mbox{.}}{2011}]%
        {gulwani2011}
\bibfield{author}{\bibinfo{person}{Sumit Gulwani}, \bibinfo{person}{Susmit
  Jha}, \bibinfo{person}{Ashish Tiwari}, {and} \bibinfo{person}{Ramarathnam
  Venkatesan}.} \bibinfo{year}{2011}\natexlab{}.
\newblock \showarticletitle{Synthesis of loop-free programs}.
\newblock \bibinfo{journal}{\emph{ACM SIGPLAN Notices}} \bibinfo{volume}{46},
  \bibinfo{number}{6} (\bibinfo{year}{2011}), \bibinfo{pages}{62--73}.
\newblock


\bibitem[\protect\citeauthoryear{Haney, Machanavajjhala, Abowd, Graham,
  Kutzbach, and Vilhuber}{Haney et~al\mbox{.}}{2017}]%
        {Haney:2017:UCF}
\bibfield{author}{\bibinfo{person}{Samuel Haney}, \bibinfo{person}{Ashwin
  Machanavajjhala}, \bibinfo{person}{John~M. Abowd}, \bibinfo{person}{Matthew
  Graham}, \bibinfo{person}{Mark Kutzbach}, {and} \bibinfo{person}{Lars
  Vilhuber}.} \bibinfo{year}{2017}\natexlab{}.
\newblock \showarticletitle{Utility Cost of Formal Privacy for Releasing
  National Employer-Employee Statistics}. In
  \bibinfo{booktitle}{\emph{Proceedings of the 2017 ACM International
  Conference on Management of Data}} (Chicago, Illinois, USA)
  \emph{(\bibinfo{series}{SIGMOD '17})}. \bibinfo{publisher}{ACM},
  \bibinfo{address}{New York, NY, USA}, \bibinfo{pages}{1339--1354}.
\newblock
\showISBNx{978-1-4503-4197-4}
\urldef\tempurl%
\url{https://doi.org/10.1145/3035918.3035940}
\showDOI{\tempurl}


\bibitem[\protect\citeauthoryear{Hunt and Sands}{Hunt and Sands}{2006}]%
        {Hunt:flowsensitive}
\bibfield{author}{\bibinfo{person}{Sebastian Hunt} {and} \bibinfo{person}{David
  Sands}.} \bibinfo{year}{2006}\natexlab{}.
\newblock \showarticletitle{On Flow-sensitive Security Types}. In
  \bibinfo{booktitle}{\emph{Conference Record of the 33rd ACM SIGPLAN-SIGACT
  Symposium on Principles of Programming Languages}} (Charleston, South
  Carolina, USA) \emph{(\bibinfo{series}{POPL '06})}. \bibinfo{publisher}{ACM},
  \bibinfo{address}{New York, NY, USA}, \bibinfo{pages}{79--90}.
\newblock
\showISBNx{1-59593-027-2}


\bibitem[\protect\citeauthoryear{Johnson, Near, and Song}{Johnson
  et~al\mbox{.}}{2018}]%
        {elasticsensitivity}
\bibfield{author}{\bibinfo{person}{Noah Johnson}, \bibinfo{person}{Joseph~P
  Near}, {and} \bibinfo{person}{Dawn Song}.} \bibinfo{year}{2018}\natexlab{}.
\newblock \showarticletitle{Towards practical differential privacy for SQL
  queries}.
\newblock \bibinfo{journal}{\emph{Proceedings of the VLDB Endowment}}
  \bibinfo{volume}{11}, \bibinfo{number}{5} (\bibinfo{year}{2018}),
  \bibinfo{pages}{526--539}.
\newblock


\bibitem[\protect\citeauthoryear{Kennedy and Eberhart}{Kennedy and
  Eberhart}{1995}]%
        {pso}
\bibfield{author}{\bibinfo{person}{J. Kennedy} {and} \bibinfo{person}{R.
  Eberhart}.} \bibinfo{year}{1995}\natexlab{}.
\newblock \showarticletitle{Particle swarm optimization}. In
  \bibinfo{booktitle}{\emph{Proceedings of ICNN'95 - International Conference
  on Neural Networks}}, Vol.~\bibinfo{volume}{4}. \bibinfo{publisher}{IEEE},
  \bibinfo{pages}{1942--1948 vol.4}.
\newblock
\urldef\tempurl%
\url{https://doi.org/10.1109/ICNN.1995.488968}
\showDOI{\tempurl}


\bibitem[\protect\citeauthoryear{Lyu, Su, and Li}{Lyu et~al\mbox{.}}{2017}]%
        {ninghuisparse}
\bibfield{author}{\bibinfo{person}{Min Lyu}, \bibinfo{person}{Dong Su}, {and}
  \bibinfo{person}{Ninghui Li}.} \bibinfo{year}{2017}\natexlab{}.
\newblock \showarticletitle{Understanding the sparse vector technique for
  differential privacy}.
\newblock \bibinfo{journal}{\emph{Proceedings of the VLDB Endowment}}
  \bibinfo{volume}{10}, \bibinfo{number}{6} (\bibinfo{year}{2017}),
  \bibinfo{pages}{637--648}.
\newblock


\bibitem[\protect\citeauthoryear{{Machanavajjhala}, {Kifer}, {Abowd}, {Gehrke},
  and {Vilhuber}}{{Machanavajjhala} et~al\mbox{.}}{2008}]%
        {ashwin08:map}
\bibfield{author}{\bibinfo{person}{A. {Machanavajjhala}}, \bibinfo{person}{D.
  {Kifer}}, \bibinfo{person}{J. {Abowd}}, \bibinfo{person}{J. {Gehrke}}, {and}
  \bibinfo{person}{L. {Vilhuber}}.} \bibinfo{year}{2008}\natexlab{}.
\newblock \showarticletitle{Privacy: Theory meets Practice on the Map}. In
  \bibinfo{booktitle}{\emph{2008 IEEE 24th International Conference on Data
  Engineering}}. \bibinfo{publisher}{IEEE}, \bibinfo{address}{Piscataway, NJ,
  USA}, \bibinfo{pages}{277--286}.
\newblock
\urldef\tempurl%
\url{https://doi.org/10.1109/ICDE.2008.4497436}
\showDOI{\tempurl}


\bibitem[\protect\citeauthoryear{McSherry}{McSherry}{2009}]%
        {pinq}
\bibfield{author}{\bibinfo{person}{Frank~D. McSherry}.}
  \bibinfo{year}{2009}\natexlab{}.
\newblock \showarticletitle{Privacy Integrated Queries: An Extensible Platform
  for Privacy-preserving Data Analysis}. In
  \bibinfo{booktitle}{\emph{Proceedings of the 2009 ACM SIGMOD International
  Conference on Management of Data}} (Providence, Rhode Island, USA)
  \emph{(\bibinfo{series}{SIGMOD '09})}. \bibinfo{publisher}{ACM},
  \bibinfo{address}{New York, NY, USA}, \bibinfo{pages}{19--30}.
\newblock
\showISBNx{978-1-60558-551-2}


\bibitem[\protect\citeauthoryear{Miranda}{Miranda}{2018}]%
        {pyswarmsJOSS2018}
\bibfield{author}{\bibinfo{person}{Lester James~V. Miranda}.}
  \bibinfo{year}{2018}\natexlab{}.
\newblock \showarticletitle{{P}y{S}warms, a research-toolkit for {P}article
  {S}warm {O}ptimization in {P}ython}.
\newblock \bibinfo{journal}{\emph{Journal of Open Source Software}}
  \bibinfo{volume}{3} (\bibinfo{year}{2018}).
\newblock
Issue 21.
\urldef\tempurl%
\url{https://doi.org/10.21105/joss.00433}
\showDOI{\tempurl}


\bibitem[\protect\citeauthoryear{Reed and Pierce}{Reed and Pierce}{2010}]%
        {Fuzz}
\bibfield{author}{\bibinfo{person}{Jason Reed} {and}
  \bibinfo{person}{Benjamin~C. Pierce}.} \bibinfo{year}{2010}\natexlab{}.
\newblock \showarticletitle{Distance Makes the Types Grow Stronger: A Calculus
  for Differential Privacy}. In \bibinfo{booktitle}{\emph{Proceedings of the
  15th ACM SIGPLAN International Conference on Functional Programming}}
  (Baltimore, Maryland, USA) \emph{(\bibinfo{series}{ICFP '10})}.
  \bibinfo{publisher}{ACM}, \bibinfo{address}{New York, NY, USA},
  \bibinfo{pages}{157--168}.
\newblock
\showISBNx{978-1-60558-794-3}
\urldef\tempurl%
\url{https://doi.org/10.1145/1863543.1863568}
\showDOI{\tempurl}


\bibitem[\protect\citeauthoryear{Roy, Hsu, and Albarghouthi}{Roy
  et~al\mbox{.}}{2021}]%
        {learning}
\bibfield{author}{\bibinfo{person}{S. Roy}, \bibinfo{person}{J. Hsu}, {and}
  \bibinfo{person}{A. Albarghouthi}.} \bibinfo{year}{2021}\natexlab{}.
\newblock \showarticletitle{Learning Differentially Private Mechanisms}. In
  \bibinfo{booktitle}{\emph{IEEE Symposium on Security and Privacy (SP)}}.
  \bibinfo{publisher}{IEEE Computer Society}, \bibinfo{address}{Los Alamitos,
  CA, USA}, \bibinfo{pages}{1033--1046}.
\newblock
\urldef\tempurl%
\url{https://doi.org/10.1109/SP40001.2021.00060}
\showDOI{\tempurl}


\bibitem[\protect\citeauthoryear{Sabelfeld and Myers}{Sabelfeld and
  Myers}{2003}]%
        {sm-jsac}
\bibfield{author}{\bibinfo{person}{Andrei Sabelfeld} {and}
  \bibinfo{person}{Andrew~C. Myers}.} \bibinfo{year}{2003}\natexlab{}.
\newblock \showarticletitle{Language-Based Information-Flow Security}.
\newblock \bibinfo{journal}{\emph{IEEE Journal on Selected Areas in
  Communications}} \bibinfo{volume}{21}, \bibinfo{number}{1}
  (\bibinfo{date}{Jan.} \bibinfo{year}{2003}), \bibinfo{pages}{5--19}.
\newblock


\bibitem[\protect\citeauthoryear{Smith and Albarghouthi}{Smith and
  Albarghouthi}{2019}]%
        {synthesisdp}
\bibfield{author}{\bibinfo{person}{Calvin Smith} {and} \bibinfo{person}{Aws
  Albarghouthi}.} \bibinfo{year}{2019}\natexlab{}.
\newblock \showarticletitle{Synthesizing Differentially Private Programs}.
\newblock \bibinfo{journal}{\emph{Proc. ACM Program. Lang.}}
  \bibinfo{volume}{3}, \bibinfo{number}{ICFP}, Article \bibinfo{articleno}{94}
  (\bibinfo{date}{July} \bibinfo{year}{2019}), \bibinfo{numpages}{29}~pages.
\newblock
\urldef\tempurl%
\url{https://doi.org/10.1145/3341698}
\showDOI{\tempurl}


\bibitem[\protect\citeauthoryear{Solar-Lezama, Tancau, Bodik, Seshia, and
  Saraswat}{Solar-Lezama et~al\mbox{.}}{2006}]%
        {CEGIS}
\bibfield{author}{\bibinfo{person}{Armando Solar-Lezama},
  \bibinfo{person}{Liviu Tancau}, \bibinfo{person}{Rastislav Bodik},
  \bibinfo{person}{Sanjit Seshia}, {and} \bibinfo{person}{Vijay Saraswat}.}
  \bibinfo{year}{2006}\natexlab{}.
\newblock \showarticletitle{Combinatorial Sketching for Finite Programs}. In
  \bibinfo{booktitle}{\emph{Proceedings of the 12th International Conference on
  Architectural Support for Programming Languages and Operating Systems}} (San
  Jose, California, USA) \emph{(\bibinfo{series}{ASPLOS XII})}.
  \bibinfo{publisher}{Association for Computing Machinery},
  \bibinfo{address}{New York, NY, USA}, \bibinfo{pages}{404–415}.
\newblock
\showISBNx{1595934510}
\urldef\tempurl%
\url{https://doi.org/10.1145/1168857.1168907}
\showDOI{\tempurl}


\bibitem[\protect\citeauthoryear{Team}{Team}{2017}]%
        {applediffp}
\bibfield{author}{\bibinfo{person}{Apple Differential~Privacy Team}.}
  \bibinfo{year}{2017}\natexlab{}.
\newblock \bibinfo{title}{Learning with Privacy at Scale}.
\newblock
\newblock
\urldef\tempurl%
\url{https://machinelearning.apple.com/2017/12/06/learning-with-privacy-at-scale.html}
\showURL{%
\tempurl}


\bibitem[\protect\citeauthoryear{Volpano, Smith, and Irvine}{Volpano
  et~al\mbox{.}}{1996}]%
        {vsi96}
\bibfield{author}{\bibinfo{person}{Dennis Volpano}, \bibinfo{person}{Geoffrey
  Smith}, {and} \bibinfo{person}{Cynthia Irvine}.}
  \bibinfo{year}{1996}\natexlab{}.
\newblock \showarticletitle{A Sound Type System for Secure Flow Analysis}.
\newblock \bibinfo{journal}{\emph{Journal of Computer Security}}
  \bibinfo{volume}{4}, \bibinfo{number}{3} (\bibinfo{year}{1996}),
  \bibinfo{pages}{167--187}.
\newblock


\bibitem[\protect\citeauthoryear{Wang, Ding, Kifer, and Zhang}{Wang
  et~al\mbox{.}}{2020}]%
        {checkdp}
\bibfield{author}{\bibinfo{person}{Yuxin Wang}, \bibinfo{person}{Zeyu Ding},
  \bibinfo{person}{Daniel Kifer}, {and} \bibinfo{person}{Danfeng Zhang}.}
  \bibinfo{year}{2020}\natexlab{}.
\newblock \showarticletitle{CheckDP: An Automated and Integrated Approach for
  Proving Differential Privacy or Finding Precise Counterexamples}. In
  \bibinfo{booktitle}{\emph{Proceedings of the 2020 ACM SIGSAC Conference on
  Computer and Communications Security}} (Virtual Event, USA)
  \emph{(\bibinfo{series}{CCS '20})}. \bibinfo{publisher}{Association for
  Computing Machinery}, \bibinfo{address}{New York, NY, USA},
  \bibinfo{pages}{919–938}.
\newblock
\showISBNx{9781450370899}
\urldef\tempurl%
\url{https://doi.org/10.1145/3372297.3417282}
\showDOI{\tempurl}


\bibitem[\protect\citeauthoryear{Wang, Ding, Wang, Kifer, and Zhang}{Wang
  et~al\mbox{.}}{2019}]%
        {shadowdp}
\bibfield{author}{\bibinfo{person}{Yuxin Wang}, \bibinfo{person}{Zeyu Ding},
  \bibinfo{person}{Guanhong Wang}, \bibinfo{person}{Daniel Kifer}, {and}
  \bibinfo{person}{Danfeng Zhang}.} \bibinfo{year}{2019}\natexlab{}.
\newblock \showarticletitle{Proving Differential Privacy with Shadow
  Execution}. In \bibinfo{booktitle}{\emph{Proceedings of the 40th ACM SIGPLAN
  Conference on Programming Language Design and Implementation}} (Phoenix, AZ,
  USA) \emph{(\bibinfo{series}{PLDI 2019})}. \bibinfo{publisher}{ACM},
  \bibinfo{address}{New York, NY, USA}, \bibinfo{pages}{655--669}.
\newblock
\showISBNx{978-1-4503-6712-7}
\urldef\tempurl%
\url{https://doi.org/10.1145/3314221.3314619}
\showDOI{\tempurl}


\bibitem[\protect\citeauthoryear{Zhang and Kifer}{Zhang and Kifer}{2017}]%
        {lightdp}
\bibfield{author}{\bibinfo{person}{Danfeng Zhang} {and} \bibinfo{person}{Daniel
  Kifer}.} \bibinfo{year}{2017}\natexlab{}.
\newblock \showarticletitle{LightDP: Towards Automating Differential Privacy
  Proofs}. In \bibinfo{booktitle}{\emph{Proceedings of the 44th ACM SIGPLAN
  Symposium on Principles of Programming Languages}} (Paris, France)
  \emph{(\bibinfo{series}{POPL 2017})}. \bibinfo{publisher}{ACM},
  \bibinfo{address}{New York, NY, USA}, \bibinfo{pages}{888--901}.
\newblock
\showISBNx{978-1-4503-4660-3}


\end{thebibliography}
